\documentclass[twocolumn]{aastex631}
\usepackage{color}
\received{}
\revised{}
\accepted{}
\submitjournal{ApJS}

\shorttitle{Neutrino oscillation with MC method}
\shortauthors{Kato et al.}

\begin{document}

\title{Neutrino transport with Monte Carlo method: II. Quantum Kinetic Equations}

\correspondingauthor{Chinami Kato}
\email{ckato@rs.tus.ac.jp}

\author{Chinami Kato}
\affiliation{Department of Science and Technology, Tokyo University of Science, 2641 Yamazaki, Noda-shi, Chiba Prefecture 278-8510, Japan}
\affiliation{Department of Aerospace Engineering, Tohoku University, 6-6-01 Aramaki-Aza-Aoba, Aoba-ku, Sendai 980-8579, Japan}

\author{Hiroki Nagakura}
\affiliation{National Astronomical Observatory of Japan, 2-21-1 Osawa, Mitaka, Tokyo 181-8588, Japan}
\affiliation{Department of Astrophysical Sciences, Princeton University, Princeton, NJ 08544}

\author{Taiki Morinaga}
\affiliation{Graduate School of Advanced Science and Engineering, Waseda University, 3-4-1 Okubo, Shinjuku, Tokyo 169-8555, Japan}

\begin{abstract}
Neutrinos have an unique quantum feature as flavor conversions. Recent studies suggested that collective neutrino oscillations play important roles in high-energy astrophysical phenomena. Quantum kinetic equation (QKE) is capable of describing the neutrino flavor conversion, transport and matter collision self-consistently. However, we have experienced many technical difficulties in their numerical implementation. In this paper, we present a new QKE solver based on Monte Carlo (MC) approach. This is an upgraded version of our classical MC neutrino transport solver; in essence, a flavor degree of freedom including mixing state is added into each MC particle. This extension requires updating numerical treatments of collision terms, in particular for scattering processes. We deal with the technical problem by generating a new MC particle at each scattering event. 
To reduce statistical noise inherent in MC methods, we develop the effective mean free path method. This suppresses a sudden change of flavor state due to collisions without increasing the number of MC particles.
We present a suite of code tests to validate these new modules with comparing to the results reported in previous studies. Our QKE-MC solver is developed with fundamentally different philosophy and design from other deterministic- and mesh methods, suggesting that it will be complementary to others, and potentially provide new insights into physical processes of neutrino dynamics.
\end{abstract}






\section{Introduction}
In hot and dense media,
neutrinos have important roles as effective carriers of energy and lepton-number.
During the advanced stages of massive star evolution ($M \gtrsim 10 M_{\sun}$), neutrino cooling dictates the thermal evolution of the stellar core \citep{farag2020,Odrzywolek:2010zz}. In the subsequent core-collapse supernova (CCSN), neutrinos are produced abundantly via various weak interaction processes. The neutrino transport and their matter interactions may be central to the explosion mechanism \citep{janka2012,kotake2012,foglizzo2015,mirizzi2016,muller2019,muller2020,burrows2021}, and of nucleosynthesis of heavy elements \citep{martinezpinedo2014,harris2017,wanajo2018,sieverding2020,wit2021}. Regardless of the detailed explosion mechanism, the emitted neutrinos from Galactic events are observable signals by current and future-planned terrestrial detectors. The direct detection of neutrinos from these nearby CCSNe will provide high statistics neutrino data, that will shed light on the inner dynamics of CCSN \citep{horiuchi2018,rosso2018,suwa2019,nagakura2021c,nagakura2021,Li2021,mori2021}. The remnant of binary neutron star merger (BNSM) is another exciting production site of neutrinos, that has received huge interest with the dawn of multi-messenger astrophysics \citep{abbott2017}. Although neutrino dynamics is a key ingredient to account for the time evolution of the system, the detailed flavor-dependent features of energy spectrum and angular distributions of neutrinos are still uncertain. Accurate treatments of neutrino transport and neutrino-matter interactions are indispensable towards understanding of the global dynamics, ejecta composition and observable signals including kilonova transients \citep{sekiguchi2015,thielemann2017,siegel2017,radice2018,miller2019,kawaguchi2021}.

There is growing experimental evidence that neutrinos' flavor states oscillate from one to another during the propagation, known as neutrino oscillation. The flavor mixing is attributed to the fact that neutrinos are massive particles, and that their mass- and flavor eigenstates are not identical each other. In a vacuum, the flavor conversion is dictated by the difference of eigenvalues of the mass eigenstates (i.e., the mass difference of neutrinos), and non-zero mixing angles.
Accurate determination of these oscillation parameters is the major challenge in experimental physics, that has made remarkable progress over the past decades \citep{Abe:2013gga,minos2014,Abe:2014ugx,SK2018,t2k2018,icecube2018,nova2019,esteban2020}.

Neutrinos experience refractive effects when they propagate in matter. This changes the dispersion relation of neutrinos, implying that flavor conversions are altered. The matter effect dominates over the vacuum oscillation in CCSNe and remnants of BNSM. This may induce strong flavor mixing at the resonance MSW matter density \citep{wolfenstein1978}. Another important consequence of the matter effect is that the mixing angles between flavor- and effective mass eigenstates are substantially reduced at a high density region, indicating that the flavor conversion is strongly suppressed. This is a rationale behind the assumption that no flavor mixing occur in a CCSN/BNSM core \citep[e.g.,][]{fuller1987,dighe2000}; indeed, almost every numerical simulation has been carried out with the assumption.

However, refractive effects by neutrino-self interactions potentially induce strong flavor conversion, also known as collective neutrino oscillations \citep{stuart1993,sigl1993,sigl1995,sawyer2005}. They are genuinely a non-linear phenomenon with large off-diagonal components of the Hamiltonian potential in the flavor-basis, suggesting that the flavor conversion is qualitatively different from that induced by the vacuum and matter potentials. The collective oscillation involves two distinct modes: fast- and slow flavor conversions. The time scales of these flavor conversions are very short, in particular for the fast mode; the oscillation frequency increases proportionally to the neutrino number density. Although the growth rate of the slow mode is relatively mild, various interesting phenomena have been found, for instances, spectral swaps/splits, bipolar and synchronized oscillations, although the occurrence of these phenomena in CCSN and BNSM environments are still being debated \citep{duan2006,Duan:2006an,dasgupta2009,tian2017,martin2020b}. 

In the last several years, there has been significant progress of our theoretical understanding of collective neutrino oscillations. Some of the progress are owed to the linear stability analysis that provides not only insights into characteristics of collective neutrino oscillations but also criteria to assess the onset of flavor conversions \citep{dasgupta2017,nagakura2019,milad2019,abbar2020,morinaga2020,milad2020}. Regarding the fast mode, electron neutrino lepton number crossings  (ELN crossings) in angular distributions are associated with the instability \citep{morinaga2021}. The recent studies suggested that ELN crossings have been frequently observed in the neutrino data of CCSN and BNSM simulations \citep{tamborra2017,wu2017b,dasgupta2018,glas2020,abbar2020b,capozzi2020,abbar2021b,abbar2021}, that increases the possibility of occurrence of fast flavor conversion in real CCSN/BNSM environments.
Concerning the slow mode, some recent studies have demonstrated that this type of flavor oscillation affects observational consequences of CCSNe such as nucleosynthesis and neutrino signals \citep{chakraborty2010,zaizen2018,sasaki2017,wu2017,george2020}. These results motivate more detailed studies of collective neutrino oscillations.

Aside from a flavor mixing, neutrinos experience absorption and momentum-exchanged scattering\footnote{We note that the term "momentum-exchanged scattering" is used to distinguish from the forward scattering generating neutrino refraction.}
during the propagation in medium. The macroscopic properties of neutrino radiation field in CCSN and BNSM strongly hinge on these collision term. They also affect the dynamics of collective neutrino oscillations; for instances, both fast and slow modes may be triggered by neutrino scatterings with nucleons or nuclei \citep{cherry2013,cherry2020,morinaga2020,zaizen2020}; the characteristics of flavor conversions may be altered by these scatterings \citep{capozzi2019,johns2020,shalgar2021b,shalgar2021c,martin2021}; \cite{johns2021} suggested very recently that a new type of instability, collisional flavor instability, appears due to collision terms. Hence, the coupling between collective neutrino oscillations and collision terms is currently of great interest. On the other hand, time scales of neutrino-matter collisions are usually longer than that of neutrino oscillations, indicating that the investigation of the interplay to neutrino-matter collisions should inevitably be in non-linear phase of collective oscillations. The direct numerical simulation is, hence, the most useful approach for the investigation.

The consistent treatment of neutrino transport, flavor conversion and neutrino-matter collisions requires the quantum kinetic description of the neutrino-radiation field. The quantum kinetic equation (QKE) provides the basic equation, that can be concisely written in the mean-field approximation (see also Eq.~\ref{eq:basiceqrho}). However, the direct numerical simulation is not an easy task, but rather a grand challenge in computational physics. This is mainly attributed to the large disparity of time- and spatial scales between flavor conversions and global (macroscopic) system. Recent studies in \citet{johns2020,sherwood2021} also found that fast flavor conversions generally involve cascade phenomena, in which fast flavor conversions trigger the energy exchange between large- to small scales. This suggests that small scale structures of both spatial- and momentum spaces need to be resolved, posing another challenge in numerical simulations. For these reasons, the currently feasible approach is either local simulations \citep{abbar2019,sherwood2019,martin2020,shalgar2021b,sherwood2021,martin2021,zaizen2021} or large-scale simulations with employing crude approximations and simplifications \citep[see, e.g.,][]{duan2006,dasgupta2008,cherry2012,zaizen2018,stapleford2020,Li2021}.

Regarding the methodologies, there are mainly two options: deterministic- and probabilistic approaches. The former has been popular in QKE neutrino transport. This may be because the numerical technique can be simply extended from classical radiation transport; indeed, it is a good compatibility with finite volume methods (but see a PIC method in \cite{sherwood2021}). On the other hand, there are no previous work that employ the probabilistic way in QKE neutrino transport. The main reason of the absence of demonstration is that probabilistic treatments are not necessary in QKE neutrino transport, unless the collision term is taken into account. As already mentioned, however, there is a growing interest in interplay between flavor conversions and collision terms, indicating that the detailed investigation of the coupling system will be made more vigorously in future. It should also be mentioned that, since all numerical simulations involve a certain level of approximation, it is necessary to assess the reliability of their results. Independent numerical approach will help the assessment, and provide confidence in the accuracy of each result. In fact, such collaborative code-comparison projects can be seen in classical neutrino transport \citep{sherwood2017,oconnor2018}.

In this paper, we present a new QKE neutrino transport solver based on a Monte Carlo (MC) method, which is a representative probabilistic approach.
This QKE-MC code is an extension version of our classical MC neutrino transport one \citep{kato2020}, in which the neutrino transport with various weak processes are well tested in the context of CCSN simulations. On the other hand, there is a non-trivial issue: how the quantum kinetic states can be handled in the particle method. Addressing this issue is one of the focuses in this paper.
We develop new modules to reduce the statistical noise inherent to the MC approach, and to increase the computational efficiency. After presenting the numerical implementation, we demonstrate a suite of code tests including fast flavor conversions with momentum-exchanged scatterings to test the capability of these new modules.

This paper is organized as follows. We start with providing an overview of the QKE for neutrino transport in Sec. \ref{ch2}.
The essential methodologies in our QKE-MC solver are described in Sec.~\ref{ch3}. Before entering into the detail of code tests, we describe the basic assumption in Sec.~\ref{sec:basicassum}. From Secs.~\ref{ch4} to~\ref{ch6}, we present code tests for vacuum, matter and collective neutrino oscillations, respectively.
Other tests regarding QKE with scatterings are demonstrated in Sec. \ref{ch7}.
We conclude the paper with a summary and discussions in Section \ref{ch8}. Throughout this paper, we use the metric signature of $(- + + +)$. Unless otherwise stated, Greek subscripts denote the spacetime components, and we work in units with $c=\hbar=1$, where $c$ and $\hbar$ denote the speed of light and reduced Planck constant, respectively.


\section{Basic equation} \label{ch2}

The QKE for neutrino transport can be written as \citep[see also][]{sigl1993,volpe2015},
\begin{eqnarray}
 && i\left(p^\alpha\left. \frac{\partial \rho}{\partial x^\alpha}\right|_{p^i} -\Gamma^i_{\alpha\beta} p^\alpha p^\beta 
    \left. \frac{\partial \rho}{\partial p^i }\right|_{x^\mu}\right) \nonumber \\
  &&\ \ \ \ \ = (- p^\mu \zeta_\mu)
    \left[H_{\rm{vac}}+H_{\rm{mat}}+H_{\nu\nu}, \rho \right] + C, \nonumber \\ 
 && i\left(p^\alpha\left. \frac{\partial \bar{\rho}}{\partial x^\alpha}\right|_{p^i} -\Gamma^i_{\alpha\beta} p^\alpha p^\beta 
    \left. \frac{\partial \bar{\rho}}{\partial p^i }\right|_{x^\mu}\right) \nonumber \\
 &&\ \ \ \ \  =(- p^\mu \zeta_\mu)
    \left[H_{\rm{vac}}^*-H_{\rm{mat}}^*-H_{\nu\nu}^*, \bar{\rho} \right] + \bar{C}, \label{eq:basiceqrho}
\end{eqnarray}
where $\rho$ $(\bar{\rho})$ is the density matrix for neutrinos (anti-neutrinos); 
$x^\mu$ denotes the space-time coordinate;
$p^{\mu}$ denotes the four-momentum of neutrinos; $C$ $(\bar{C})$ is the collision term for neutrinos (anti-neutrinos); $\Gamma$ is the Christoffel symbols; $\zeta^{\mu}$ denotes the unit vector normal to the spatial hypersurface of $t={\rm const}$, where $t$ is the laboratory time. In the expression, Hamiltonian operators are measured in the laboratory frame\footnote{We note that it can be measured with arbitrary time. In this case, we replace $(- p^\mu \zeta_\mu)$ by $(- p^\mu \zeta^{\prime}_\mu)$ in Eq.~\ref{eq:basiceqrho}, where $\zeta^{\prime}_{\mu}$ denotes the arbitrary timelike vector.}. Each potential can be given as
\begin{eqnarray}
   H_{\rm{vac}} &=& U\frac{M^2}{2E_\nu}U^\dagger, \nonumber \\
   H_{\rm{mat}} &=& \sqrt{2}G_F \frac{p_\mu}{E_\nu}{\rm diag}\left[j^\mu_l-\bar{j}^\mu_l\right],
   \nonumber \\
   H_{\nu\nu} &=& \sqrt{2}G_F
      \int dV_{p^\prime}
      \frac{p^\mu p_\mu^\prime}{E_\nu E_\nu^\prime}
      \left(\rho^\prime - \bar{\rho}^{\prime \ast} \right).
    \label{eq:hamiltonian}
\end{eqnarray}
In the expression, $M^2$ can be written as $M^2_{ij} = m^2_{a}\delta_{ij}$ (the index $a$ specifies the mass eigenstate of neutrinos) in the relativistic limit, where $m_a$ denote the rest-mass of $\nu_a$. $E_{\nu}=-p^\mu\zeta_\mu$ represents the energy of neutrinos measured by an observer of $\zeta_\mu$. $U$ is the unitary matrix describing the mixing between flavor and mass bases. 
$G_F$ is the Fermi coupling constant. $j^\mu_l$ represents the number current of charged-leptons, which can be written as $j^\mu_l=n_l u^\mu_l$, where $n_l$ and $u^{\mu}_l$ denotes the the number density (measured at each lepton's rest frame) and four-velocity of each charged-leptons, respectively.
$dV_p$ denotes the momentum volume element for neutrino momentum space, which can be written as
\begin{eqnarray}
dV_p &=& \frac{1}{ (2\pi)^3 } \hspace{0.5mm} d \Omega \hspace{0.5mm} d\left(\frac{E_{\nu}^3}{3}\right),
\end{eqnarray}
where $d \Omega$ denotes the solid angle of neutrinos measured in $\zeta_{\mu}$ observer\footnote{We note that $\zeta_{\mu}$ can be replaced any time-like unit vector. This depends on the choice of tetrad basis to describe the momentum space of neutrinos \cite[see e.g.,][for more detail]{shibata2014}.}.


\section{Numerical Implementations} \label{ch3}

Many modules in our QKE-MC solver are imported from classical one \citep{kato2020}. However, the flavor states in MC particles are no longer constant during the flight due to flavor conversions. A possible solution is that a flavor degree of freedom is added in each MC particle. This change also requires an update of treatments for collision terms. In this section, we describe our new ideas of the formulation and numerical techniques for them.

\subsection{Transport and oscillation} \label{subsec:transOsc}

Each MC particle of position and momentum state follows the geodesic equation during propagation,
\begin{eqnarray}
&& \frac{d x^{\mu}}{d \lambda} = p^{\mu}, \nonumber \\
&& \frac{d p^{\mu}}{d \lambda} - \Gamma^{\mu}_{\hphantom{\mu}\alpha\beta} p^{\alpha} p^{\beta} = 0, \label{eq:geo}
\end{eqnarray}
where $\lambda$ denotes the affine parameter for a trajectory of each particle. We assume neutrinos as massless particles, i.e., Eq.~\ref{eq:geo} is null geodesics.
It is a reasonable assumption for CCSN and BNSM system, since the typical energy of neutrinos are above MeV and it is much larger than the rest mass of neutrinos ($\lesssim 0.1$ eV) \citep{katrin2019}. We note that neutrinos are treated as massive particles in the evolution of flavor state (see below).

Different from conventional MC neutrino transport, the flavor degree of freedom is embedded in each MC particle, that allows us to treat the neutrino flavor conversion self-consistently. More specifically, we define the particle-matrix, $q$, on each MC particle.
The matrix structure of $q$ is the same as that of density matrix ($\rho$), i.e., $N_{f} \times N_{f}$ matrix ($N_{f}$ denotes the number of flavor). The matrix element represents the number of particles in the flavor basis, i.e., this is a natural extension from "sample particles" used in classical MC method. The density matrix ($\rho$) at each time can be computed from $q$ as
\begin{eqnarray}
\rho (\bold{x},\bold{p}) &=& \lim_{ \Delta V_x \to 0} \lim_{ \Delta V_p \to 0} \frac{1}{ \Delta V_x \hspace{0.5mm} \Delta V_p } \nonumber \\
&& \times \sum_{s} q_s \int_{\bold{x} - 0.5 \bold{\Delta{x}}}^{\bold{x} + 0.5 \bold{\Delta{x}}}
\int_{\bold{p} - 0.5 \bold{\Delta{p}}}^{\bold{p} + 0.5 \bold{\Delta{p}}} dV_x^{\prime} dV_p^{\prime} 
\hspace{0.5mm} \nonumber \\
&& \times\ \delta^3 \left(\bold{x}^{\prime}-\bold{x_{s}}(t)\right) \hspace{0.5mm} \nonumber\\
&& \times \left(2\pi\right)^3 \delta^3 \left(\bold{p}^{\prime}-\bold{p_{s}}(t)\right), \label{eq:rhofromq}
\end{eqnarray}
where $s$ denotes the index of a MC particle, and $\bar{\rho}$ (anti-neutrinos) can be computed by replacing $q \rightarrow \bar{q}$.
In the expression, $x_s$ and $p_s$ are the spatial position and momentum of $s$-th QKE-MC particle, respectively; $dV_x$ denotes a spatial three-dimensional volume element,
\begin{eqnarray}
dV_x &=& \sqrt{-g} \hspace{0.5mm} dx_1 \hspace{0.5mm} dx_2 \hspace{0.5mm} dx_3,
 \label{eq:defvolumeeleme}
\end{eqnarray}
where $g$ denotes the determinant of the space-time metric.

In practice, $\rho$ is computed by setting finite but small values of $\Delta V_x$ and $\Delta V_p$ in Eq.~\ref{eq:rhofromq}. Here we take the on-shell condition for neutrinos. The number density of neutrinos (measured in the laboratory frame) can be computed through $\rho$,
\begin{eqnarray}
  n_{ij} = \int \rho_{ij} dV_p^\prime,
\end{eqnarray}
where $i$ and $j$ specify the flavor state. The anti-neutrino counterpart can be given by replacing $\rho \rightarrow \bar{\rho}$.

We evolve $q$ and $\bar{q}$ as
\begin{eqnarray}
 i\frac{d q_{s}}{d \lambda} &=&
  (- p^\mu n_\mu )
 [H_{\rm{vac}}+H_{\rm{mat}}+H_{\rm{\nu\nu}},q_{s}], \nonumber \\
 i\frac{d \bar{q}_{s}}{d \lambda} &=&
  (- p^\mu n_\mu )
 [H_{\rm{vac}}^\ast-H_{\rm{mat}}^\ast-H_{\rm{\nu\nu}}^\ast,\bar{q}_{s}], \label{eq:basiceqforq}
\end{eqnarray}
where $d/d\lambda$ in Eq.~\ref{eq:basiceqforq} is an ordinary differentiation, which is solved with forth-order explicit Runge-Kutta method. We note that the self-interaction potential, $H_{\rm{\nu\nu}}$, can be computed from $q$ and $\bar{q}$ by using Eqs.~\ref{eq:hamiltonian}~and~\ref{eq:rhofromq}.

\subsection{Collision term} \label{subsec:col}

\begin{figure}
    \centering
    \includegraphics[width=\columnwidth]{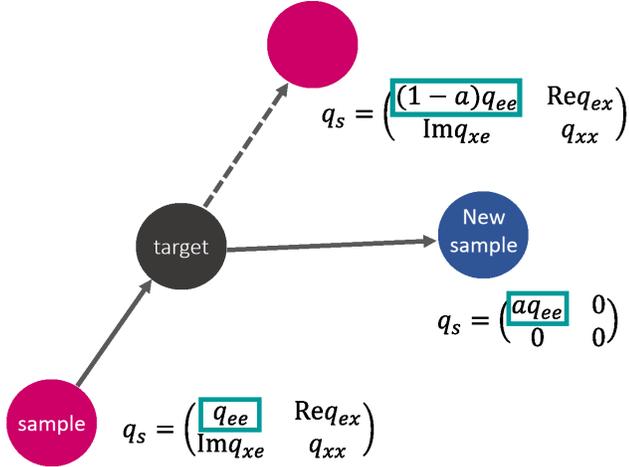}
    \caption{The schematic picture illustrating our numerical treatments of scattering process.
    As an example, we display the case that $\nu_e$ ($ee$ component of $q$) experiences scatterings. We note that $a$ corresponds to a control parameter relevant to effective mean free path (EMFP) method. See text for more detail.}
    \label{fig:overview_scat}
\end{figure}

The implementation of collision terms is well established in classical MC method, and its extension to QKE is straightforward. Below, we first describe the essence of the extension.

Let us start with emission processes. Although the neutrino matter interaction needs to be changed with the extension of QKE \citep{vlasenko2014,blaschke2016}, the implementation is the same as that used in classical MC transport. The blocking factor can also be taken into account by the same prescription used in our classical MC method. We generate new MC particles with specifying the neutrino energy and flight direction. Following the neutrino emissivity, we insert the value of neutrino number of production into each matrix element of $q_s$ (and $\bar{q}_s$) of newly generated MC particles.

For absorption process, we probabilistically determine the absorption length, the distance to a position where the neutrinos experience absorption; the procedure is exactly the same as that used in our classical MC transport. We note, however, that the length can not be assigned into each MC particle, since the reaction rate in general varies with the neutrino flavor. Therefore, it is assigned to each flavor state of MC particles. MC particles are evolved following the geodesic equation until the travel distance reaches the absorption length. At the absorption point, the corresponding matrix element of $q$ set to be zero. If $q$ becomes completely null due to the absorption, we exclude the MC particle from our simulations.

The scattering process can be handled as the similar procedure as that used in neutrino absorptions.
The scattering length, a distance to a point where the MC particle scatters with matter, is determined probabilistically. Since it is varied with neutrino flavor state, the scattering length is assigned to each matrix element of MC particle. At the scattering point, 
we generate a new MC particle, in which the neutrino energy and flight direction is chosen so as to coincide with the scattered neutrinos\footnote{These momentum space values, i.e., $p^{\mu}$, can be determined through the same probabilistic way used in our classical MC method (see \cite{kato2020}).}; i.e., it plays a role of the particle after the scattering (see Fig.~\ref{fig:overview_scat}). The matrix element of $q$ in the new MC particle is copied from that in the original particle, meanwhile we set zero in the matrix component of $q$ of the original MC particle (but see below for a corrected treatment to reduce the statistical noise). It should be emphasized that this prescription guarantees the exact conservation of neutrino numbers.

Although these treatments are capable of providing accurate solutions in principle, the statistical noise inherent in MC methods is a concern. It would affect self-interaction potentials, and then unphysical flavor oscillations may be induced. Indeed, it potentially generate artificial ELN crossings, that would kick in artificial fast neutrino flavor conversions. It should be stressed that the problem is unavoidable as long as the number of MC particle is finite\footnote{In any particle-based methods, we can always take an infinitesimal solid angle so that the number of particle is zero, indicating that artificial ELN crossings are ubiquitous at the small scale.}, indicating that the special treatment is mandatory.

To overcome the difficulty, we adopt a smoothing prescription in evaluating the self-interaction potential. Similar as mesh methods, we discretize the 6-dimensional phase space, and compute $\rho$ on each mesh by using Eq.~\ref{eq:rhofromq}. The momentum integral of neutrinos in self-interaction potential is calculated by using $\rho$. More specifically, we adopt the values of energy and flight directions at the cell center value of the momentum space grid in computations of the self-interaction potential. Although the prescription smears out a small scale structure of neutrino momentum space, the statistical noise can be controlled by changing the mesh resolution. This should be set appropriately depending on the problem.

We develop another prescription of collision terms to further reduce the statistical noise, namely effective mean free path method (hereafter we refer to it as EMFP method). In general, large statistical noise potentially appears when the MC particle experiences an absorption or scattering with matter. This is attributed to the fact that a MC particle represents a bundle of neutrinos, and that the physical states of all neutrinos change suddenly at each absorption/scattering point. Although the most straightforward way to reduce the noise is to increase the number of MC particles or to run many simulations for the increase of statistics, we propose another way to address the issue; the essence is as follows. We introduce a control parameter, $a$ ($0 \leq a \leq 1$), that determines the rate of change in each MC particle by neutrino matter interactions. Let us provide an example how we use $a$ for scattering processes. At the scattering point, $a\times q_{ij}$ (where $i, j$ specify the flavor of scattered neutrinos) is transferred from the original MC particle to the new one (see Fig.~\ref{fig:overview_scat}), i.e., the change of $q$ is softened by $a$. To compensate this effect, we reduce the scattering length by the same factor of $a$. As we show in Sec.~\ref{ch7-2}, this does not affect the physical solution of QKE system. The merit of this prescription is that the statistical noise can be reduced without pressing memory capacity in simulations.

One may wonder if the above prescription causes increasing numerical costs and makes the simulation unfeasible. In fact, the number of scattering events increases proportional to $1/a$. It should be noted, however, that this problem is under control; indeed $a=1$ recovers the original treatment. Importantly, this prescription is suited for problems of our interest: interplay between collision terms and collective neutrino oscillations. The reason is as follows. The time scale of collective neutrino oscillation is usually much shorter than that of neutrino-matter collisions, indicating that the numerical cost is dominated by transport- and oscillation terms. Indeed, for the case with $a=1$, almost every MC particle does not interact with matter during each time step ($\Delta t$) of simulations. By introducing $a$, we can manage the balance of computational time among different terms. As a result, we successfully increase the statistics without both pressing memories and computational cost. 

We make another important remark regarding the number of MC particles. In our treatments of collision term, the number of MC particles tends to increase with time, since new MC particles are generated in not only emission but also scattering processes. We address this issue by the following prescription. When the particle number almost exceeds the limit of computational resources, we combine adjacent MC particles so that the number, energy, and moment of neutrinos are conserved. Although the prescription violates the causality, this effect is expected to be minor\footnote{Regarding the causality, the computation of self-interaction potential is also an issue, since it is computed by collecting MC particles located at different spatial locations. However, the error can be estimated by the convergence study; indeed, the solution approaches to the real by increasing the number of MC particles and decreasing the phase space volume in the computation of $\rho$.}.

\subsection{Flow chart} \label{subsec:flowchart}

\begin{figure}
    \centering
    \includegraphics[width=\columnwidth]{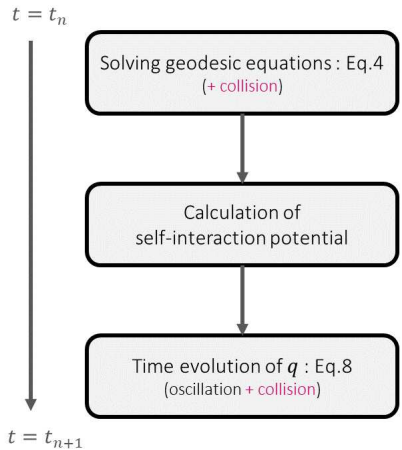}
    \caption{The flow chart for our QKE-MC solver. We highlight the parts associated with neutrino-matter interactions as purple color.}
    \label{fig:flowchart}
\end{figure}

To facilitate readers' understanding, we provide a flow chart of our QKE-MC solver in Fig.~\ref{fig:flowchart}. It summarizes the overall procedure for the time evolution of $t=t_n$ to $t_{n+1}$. We first evolve the MC particles by solving Eq.~\ref{eq:geo}. If the scattering and absorption processes are included, we change the flight direction and energy of neutrinos during the time step. The spatial transport of generated MC particles associated with scatterings (see Sec.~\ref{subsec:col}) are also handled there. It should be mentioned that we do not evolve $q$ of each MC particle at this step.

As the next step, we compute the self-interaction potential from MC particles. As mentioned in Sec.~\ref{subsec:col}, we employ a smoothing prescription to evaluate it if collision terms are taken into account. We then evolve $q$ and $\bar{q}$ by solving Eq.~\ref{eq:basiceqforq}.

There are two important remarks in the above procedure. First, we treat the time evolution of $q$ as a similar procedure of operating-splitting method; in which neutrino oscillations and collision terms are handled separately. As we have described in Sec.~\ref{subsec:col}, matrix elements of scattered MC particles needs to be corrected, and the new generated one takes over them (see also Fig.~\ref{fig:overview_scat}). This computation is done after solving Eq.~\ref{eq:basiceqforq}, i.e., completing the transport- and oscillation processes. As long as $\Delta t$ is small enough to resolve the system, the artifact due to the prescription is negligible. Second, the actual procedure of our QKE-MC solver is a bit more complicated than that described above. This is because the high-order time integration (fourth-order Runge-Kutta method) is implemented in our QKE-MC. We update the self-interaction potential at each sub-step. We note that the high-order precision is necessary only for solving Eq.~\ref{eq:basiceqforq}, i.e., to determine the evolution of neutrino flavor conversions. Except for the computation of flavor conversion, we adopt the Eulerian integration in their time integration.


\section{Basic assumptions in code tests} \label{sec:basicassum}
Although our QKE-MC solver is capable of handling the problems with arbitrary number of neutrino flavors, we assume a two flavor system throughout the code tests.
In the approximation, the density matrix has four independent components that can be written as,
\begin{eqnarray}
    \rho = \left(
    \begin{array}{cc}
      \rho_{ee} & \rho_{ex} \\
      \rho_{ex}^* & \rho_{xx}
    \end{array}
  \right), \hspace{2mm}
  \bar{\rho} = \left(
    \begin{array}{cc}
      \bar{\rho}_{ee} & \bar{\rho}_{ex} \\
      \bar{\rho}_{ex}^* & \bar{\rho}_{xx}
    \end{array}
  \right).
\end{eqnarray}

In the two flavor system, the expression with the polarization vector $\bold{P}$ and $\bar{\bold{P}}$ in the flavor space is useful. It is defined as,
\begin{eqnarray}
   \rho &=& \frac{1}{2}\left(\rho_{ee}+\rho_{xx}\right)I +  \frac{1}{2}\bold{P}\cdot \bold{\sigma}, \nonumber \\ 
   \bar{\rho} &=&
   \frac{1}{2}\left(\bar{\rho}_{ee}+\bar{\rho}_{xx}\right)I +
   \frac{1}{2}\bold{\bar{P}}\cdot \bold{\sigma^*},
\end{eqnarray}
where $\sigma$ and $I$ denotes the Pauli matrix and Unit matrix.
The component of polarization vector can be written as,
\begin{eqnarray}
   \bold{P} =
   \left(
    \begin{array}{c}
      2{\rm Re}{\rho_{ex}} \\
      -2{\rm Im}{\rho_{ex}} \\
      \rho_{ee}-\rho_{xx}
    \end{array}
  \right), \ \ 
   \bold{\bar{P}} =
    \left(
    \begin{array}{c}
      2{\rm Re}{\bar{\rho}_{ex}} \\
      2{\rm Im}{\bar{\rho}_{ex}} \\
      \bar{\rho}_{ee}-\bar{\rho}_{xx}
    \end{array}
  \right). \ \ \ \ 
  \label{polarization}
\end{eqnarray}
Using the poralization vectors, Eq.~\ref{eq:basiceqrho} can be written as Bloch-type equations:
\begin{eqnarray}
 && \frac{D }{d \lambda} 
 \bold{P} = \left[\omega \bold{B} + \sqrt{2}G_F\left(n_e-\bar{n}_e-n_x+\bar{n}_x\right) \bold{u} \right. \nonumber \\
 && \ \ \ \ \ \ \ \ \left. 
 + \sqrt{2}G_F\int dV_{p^\prime}
 \frac{p^\mu p_\mu^\prime}{E_\nu E_\nu^\prime}
 \left(\bold{P}^{\prime} - \bar{\bold{P}}^{\prime} \right) \right] \times \bold{P}, \nonumber \\
 &&\frac{D }{d \lambda}  \bold{\bar{P}} = \left[-\omega \bold{B} +  \sqrt{2}G_F\left(n_e-\bar{n}_e-n_x+\bar{n}_x\right) \bold{u}\right. \nonumber \\
 && \ \ \ \ \ \ \ \ \left. + \sqrt{2}G_F\int 
 dV_{p^\prime}
 \frac{p^\mu p_\mu^\prime}{E_\nu E_\nu^\prime}
 \left(\bold{P}^{\prime} - \bar{\bold{P}}^{\prime} \right) \right] \times \bold{\bar{P}}, \ \ \ \ \ 
 \label{eq:polarization}
\end{eqnarray}
where $\theta$ is the angle between the propagation directions of two neutrinos and $\omega \equiv \Delta m^2/2E_\nu$ is the vacuum oscillation frequency with the mass difference $\Delta m^2\equiv m_2^2 - m_1^2$. In the expression, we use the following expressions for Hamiltonian potentials,
\begin{eqnarray}
  &&H_{\rm{vac}} = \frac{1}{2}\omega\bold{B}\cdot\bold{\sigma}, \nonumber \\
  &&H_{\rm{mat}} = \frac{\sqrt{2}}{2}G_F\left(n_e-\bar{n}_e-n_x+\bar{n}_x \right) \bold{u}\cdot \bold{\sigma}, \nonumber \\
  &&H_{\nu\nu} = \frac{\sqrt{2}}{2}G_F\int dV_{p^\prime} 
  \frac{p^\mu p_\mu^\prime}{E_\nu E_\nu^\prime}
  \left(\bold{P}^{\prime}-\bold{\bar{P}}^{\prime} \right)\cdot \sigma,
\end{eqnarray}
with $\bold{u}=(0,0,1)$ and $\bold{B} = \left(\sin{2\theta_0},0,-\cos{2\theta_0}\right)$. $\theta_0$ denotes the mixing angle.
In this expression, we subtract the trace part, since they do not affect flavor conversions.

In the following code tests, we take the flat space-time.
We also assume that the matter distribution is static and homogeneous. It should be mentioned that neutrino distributions are assumed to be homogeneous throughout our code tests. More detailed tests and scientific simulations including neutrino transport under inhomogeneous backgrounds are postponed to future work.


\section{Vacuum oscillations} \label{ch4}

We start with tests of vacuum oscillations. By virtue of the linear system, we can solve the time evolution of $\rho$ analytically. They can be written in the flavor-basis as,
\begin{eqnarray}
 && \rho_{ee} = \left[1-\sin^2{2\theta_0}\sin^2{ \frac{\omega t}{2} } \right] \rho_{ee}^0 \nonumber \\
 &&\ \ \ \ \ + \sin^2{2\theta_0}\sin^2{ \frac{\omega t}{2} } \rho_{xx}^0  \nonumber \\
 &&\ \ \ \ \ - 2 \cos{2\theta_0}\sin{2\theta_0}\sin^2{ \frac{\omega t}{2} }{\rm Re}\rho_{ex}^0 \nonumber \\
 &&\ \ \ \ \ - \sin{2\theta_0}\sin{ \omega t}{\rm Im}\rho_{ex}^0, \label{eq:vac_ana_ini} \\
 && \rho_{xx} =
 \sin^2{2\theta_0}\sin^2{ \frac{\omega t}{2}} \rho_{ee}^0  \nonumber \\
 &&\ \ \ \ \ + \left[1-\sin^2{2\theta_0}\sin^2{ \frac{\omega t}{2}} \right] \rho_{xx}^0 \nonumber \\
 &&\ \ \ \ \ + 2 \cos{2\theta_0}\sin{2\theta_0}\sin^2{ \frac{\omega t}{2} }{\rm Re}\rho_{ex}^0 \nonumber \\
 &&\ \ \ \ \ + \sin{2\theta_0}\sin{  \omega t}{\rm Im}\rho_{ex}^0, \\
 &&{\rm Re}\rho_{ex} =
 \cos{2\theta_0}\sin{2\theta_0}\sin^2{ \frac{\omega t}{2}}\left(\rho_{xx}^0-\rho_{ee}^0\right) \nonumber \\
 &&\ \ \ \ \ + \left[ 1 - 2\cos^2{2\theta_0}\sin^2{ \frac{\omega t}{2}}\right] {\rm Re}\rho_{ex}^0 \nonumber \\
 &&\ \ \ \ \ - \cos{2\theta_0}\sin{ \omega t}{\rm Im}\rho_{ex}^0,\\
 &&{\rm Im}\rho_{ex} =
 \frac{1}{2} \sin{2\theta_0}\sin{ \omega t }
 \left(\rho_{ee}^0-\rho_{xx}^0\right) \nonumber \\
 &&\ \ \ \ \ + \cos{2\theta_0}\sin{ \omega t}{\rm Re}\rho_{ex}^0  \nonumber \\
 &&\ \ \ \ \ + \left[ 1 - 2\sin^2{\frac{\omega t}{2}}\right] {\rm Im}\rho_{ex}^0, \label{eq:vac_ana_fin}
\end{eqnarray}
where $\rho_{ij}^0$ denotes the initial density matrices of neutrinos (the $i$ and $j$ indexes take either $e$ or $x$).
 
In this test, we adopt the following mass and mixing parameters: $\Delta m^2 = (-)2.45\times10^{-15}\ \rm{MeV}^2$ and $\sin^2{\theta_0} = 2.24\times10^{-2}\ (2.26\times10^{-2})$ for the normal mass ordering (the inverted mass ordering) \citep{pdg14}. Hereafter, we refer to each mass hierarchy as NO and IO, respectively. The neutrino energy ($E_\nu$) is assumed to be $20$ MeV. As an initial condition, we emit a single MC particle (the flight direction can be arbitrary chosen) with setting a purely $\nu_e$ state, i.e., the $ee$ component of $q$ is 1 and others are set to be zero\footnote{Since it is a linear system (see also Eq.~\ref{eq:basiceqrho}~or~\ref{eq:basiceqforq}), the result is scale free, indicating that an arbitrary positive value can be chosen.}. During the evolution, time step ($\Delta t$) is constant with $\omega \Delta t/2 \pi = 10^{-3}$.

\begin{figure*}[htbp]
    \centering
    \hspace*{-2cm}
    \includegraphics[width=\columnwidth]{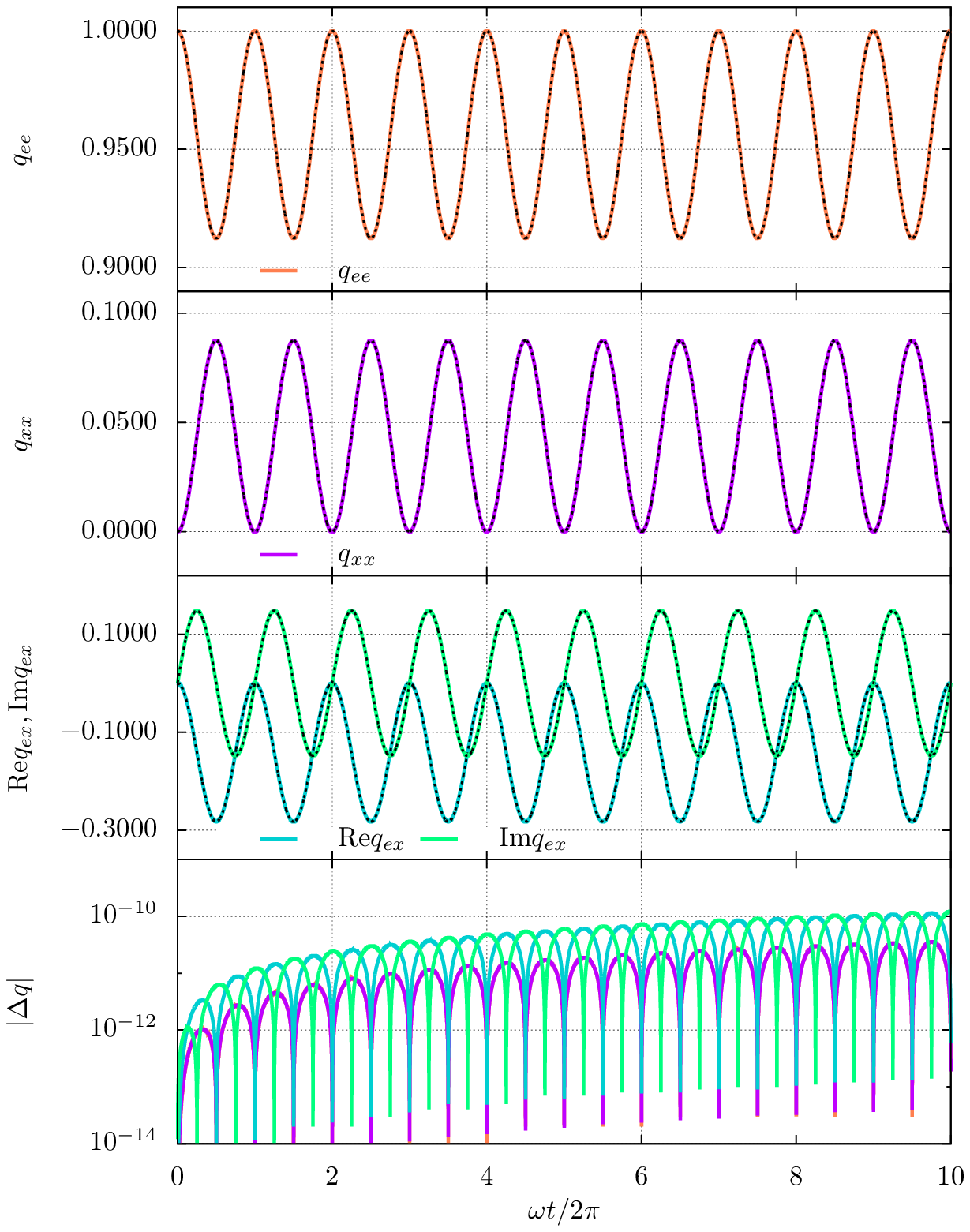}
 \includegraphics[width=\columnwidth]{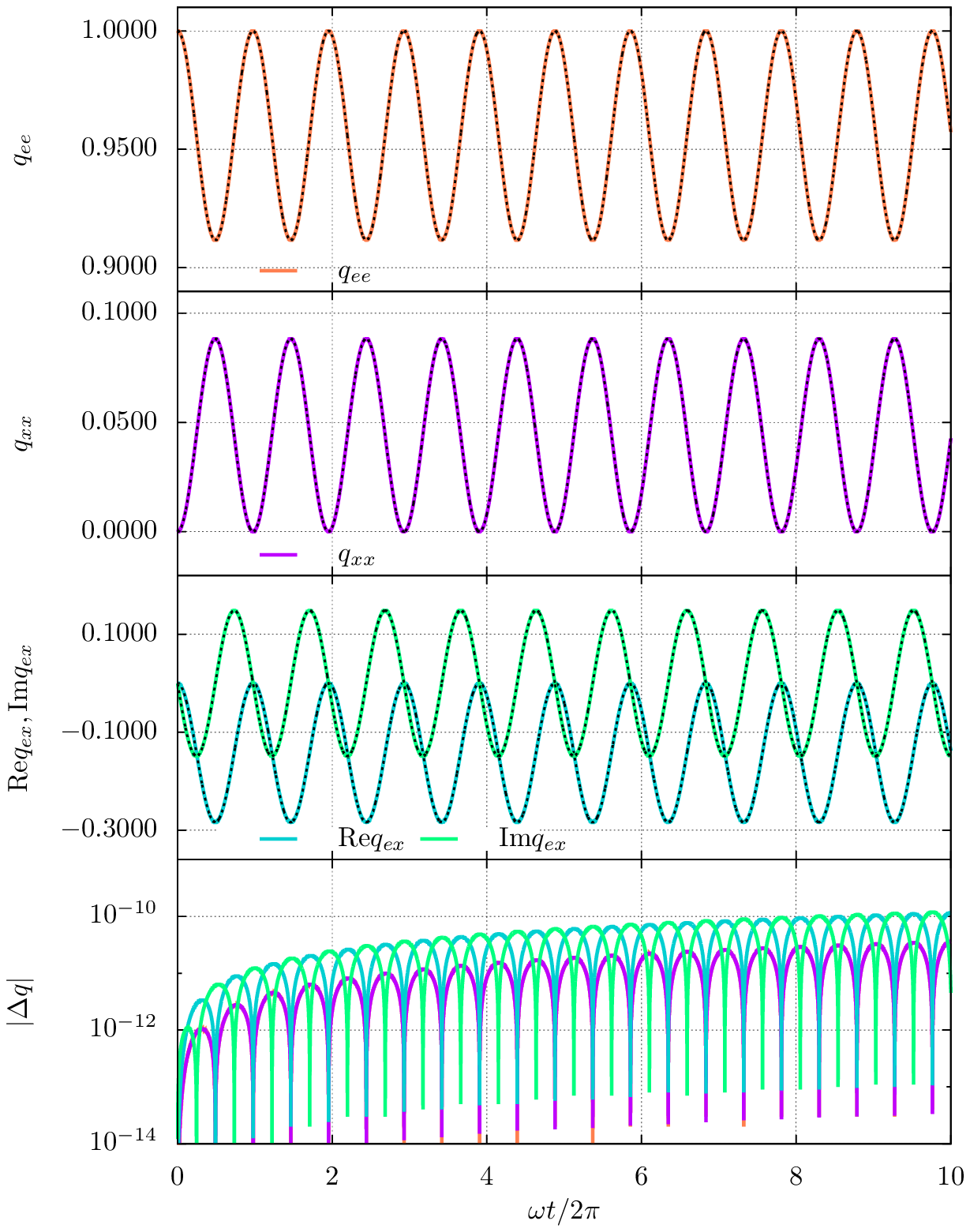}
    \caption{The time evolution of each matrix element of $q_s$ in the test of vacuum oscillations. 
    Left and right panels show the normal and inverted mass ordering, respectively.
    Solid lines represented numerical solutions and dashed black lines are analytical solutions. In the bottom panel, the errors from the analytical solutions are displayed.} 
    \label{vac_NH}
\end{figure*}

Figures~\ref{vac_NH} show the time evolution of each matrix element of $q$ in cases with NO (left panel) and IO (right panel).
The QKE-MC results are shown in solid lines while the analytical solutions are in black dashed lines.
In the bottom figures, the errors from analytic solutions are displayed. As shown in these figures, QKE-MC results are good agreement with analytic solutions for both mass hierarchies.

\begin{figure}[htbp]
    \centering
    \hspace*{-1cm}
    \includegraphics[width = \columnwidth]{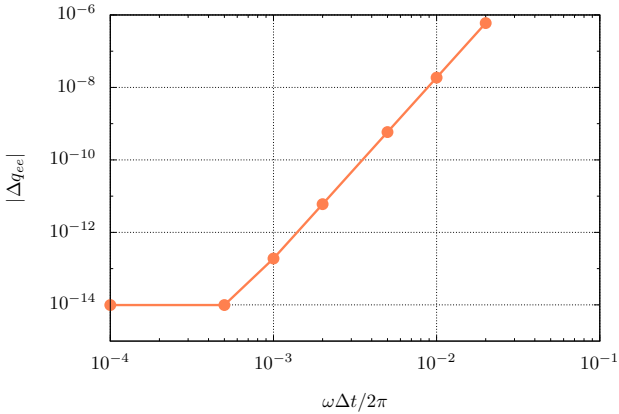}
    \caption{Numerical error as a function of $\Delta t$ for tests of vacuum oscillation in the case of the normal mass ordering (NO).}
    \label{vac_error}
\end{figure}

Figure~\ref{vac_error} portrays the convergence of our code with respect to $\Delta t$. In this check, we compare $q_{ee}$ at the time of $\omega t/2 \pi = 10$ between the numerical (QKE-MC) and analytical solutions. We confirm that our code achieved a 4th-order convergence. We note that, the round-off-error is seen in the case with $\omega \Delta t/2 \pi = 10^{-4}$, which results in degrading the convergence.

\begin{figure}[htbp]
    \centering
    \hspace*{-1.2cm}
    \includegraphics[width = \columnwidth]{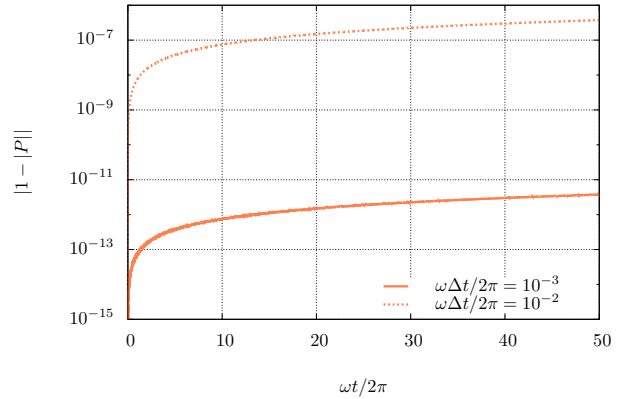}
    \caption{The time evolution of $|1-|P||$ for the test of vacuum oscillation. We show the results for simulations with two different time resolutions: $\omega \Delta t/2 \pi = 10^{-3}$ (solid line) and $10^{-2}$ (dashed line). In this figure, we assume NO for mass hierarchy.}
    \label{vac_conserve}
\end{figure}

The norm of polarization vector, $|P|$, is a conserved quantity in this test; hence, it is also useful to measure the accuracy of our code. 
In Fig.~\ref{vac_conserve}, we show the time evolution of $|1-|P||$ (we note that $|P|$ is 1 at the initial condition) for two different time step ($\omega \Delta t/2\pi = 10^{-2}$ and $10^{-3}$). We display the results with NO oscillation in the figure. This result suggests that $|P|$ is well conserved in the test, and that the accuracy is improved with decreasing $\Delta t$.

\section{Neutrino oscillation in matter} \label{ch5}

As long as the matter potential, $H_{\rm mat}$, is static and homogeneous, we can obtain exact solutions of QKE (with respect to the time evolution of the system) with vacuum and matter potentials. If only electrons surround neutrinos, the exact solution can be obtained by replacing $\theta_0\rightarrow \theta_{m}$ and $\omega \rightarrow \omega_m\equiv \Delta m_m^2/2E_\nu$ in Eqs.~\ref{eq:vac_ana_ini}-\ref{eq:vac_ana_fin}, where $\theta_m$ and $\Delta m_m^2$ can be written as,
\begin{eqnarray}
\sin{2\theta_m} &=& \frac{\sin{2\theta_0}}{\sqrt{\left(\cos{2\theta_0}\pm A\right)^2 + \sin^2{2\theta_0}}}, \label{eq:msw_sin}\\
\Delta m_m^2 &=& \Delta m^2 \sqrt{\left(\cos{2\theta_0}\pm A\right)^2+\sin^2{2\theta_0}}, \label{eq:msw_deltams}
\end{eqnarray}
with $A=2\sqrt{2}G_Fn_eE_\nu/\Delta m^2$, where $n_e$ denotes the number density of electrons.
Regarding the sign in front of $A$ in Eqs.~\ref{eq:msw_sin}~and~\ref{eq:msw_deltams}, positive and negative signs are taken in the cases of neutrinos and anti-neutrinos, respectively.
As shown in Eq.~\ref{eq:msw_sin}, the large mixing (resonance) occurs when the denominator of right hand side of Eq.~\ref{eq:msw_sin} is zero. This is equivalent to the condition of $n_e = (-)n_{e0}$ for (anti-)neutrinos, where
\begin{eqnarray}
n_{e0} = \frac{\Delta m^2\cos{2\theta_0}}{2\sqrt{2}G_FE_\nu}.
\end{eqnarray}

\begin{figure}[htbp]
    \centering
    \includegraphics[width = \columnwidth]{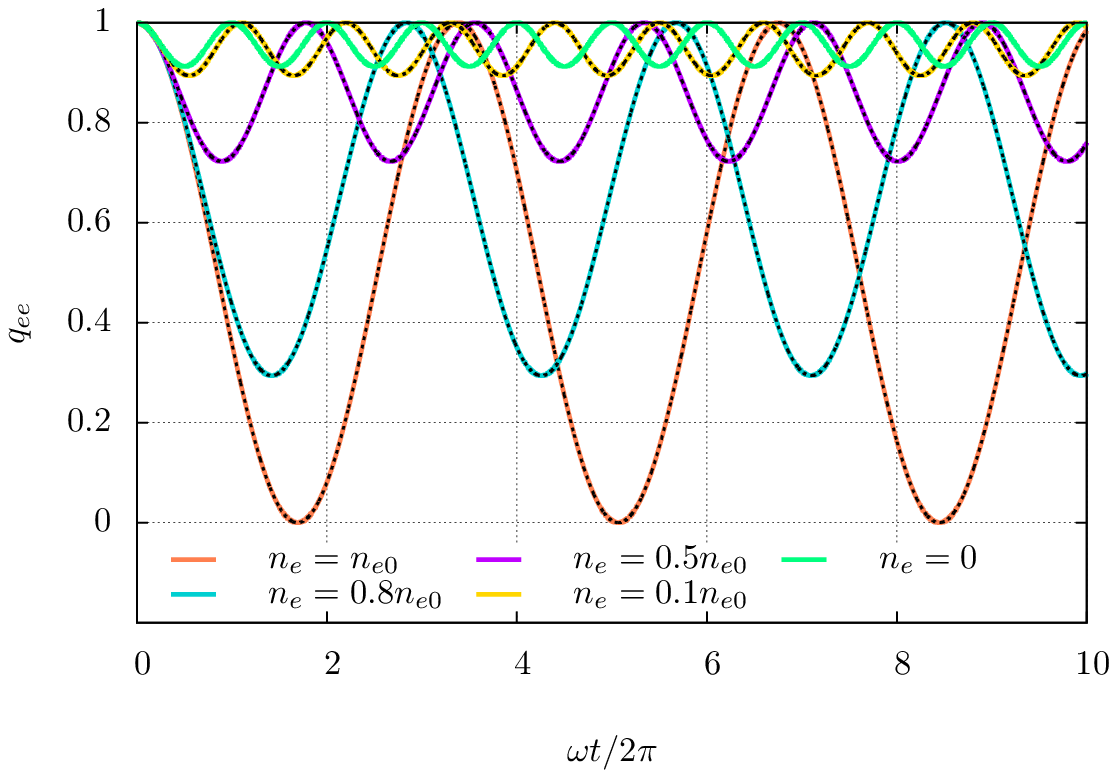}
    \caption{We show the time evolution of $\nu_e$ for the test of MSW neutrino oscillation.
    Color lines show the results for different $n_e$, and the black dashed lines represent the analytic solutions}. In this figure, we assume NO for mass hierarchy.
    \label{effect_ne}
\end{figure}

In Fig.~\ref{effect_ne}, we show the time evolution of $q_{ee}$ for various cases of $n_e$. In these tests, the initial condition of neutrinos is the same as that used in the test of vacuum oscillation (see Sec.~\ref{ch4}). The result of analytic solutions are plotted with black dashed lines. This figure suggests that our code accurately demonstrates the neutrino oscillation in matter.

\begin{figure*}[htbp]
    \centering
    \hspace*{-2cm}
    \includegraphics[width=\columnwidth]{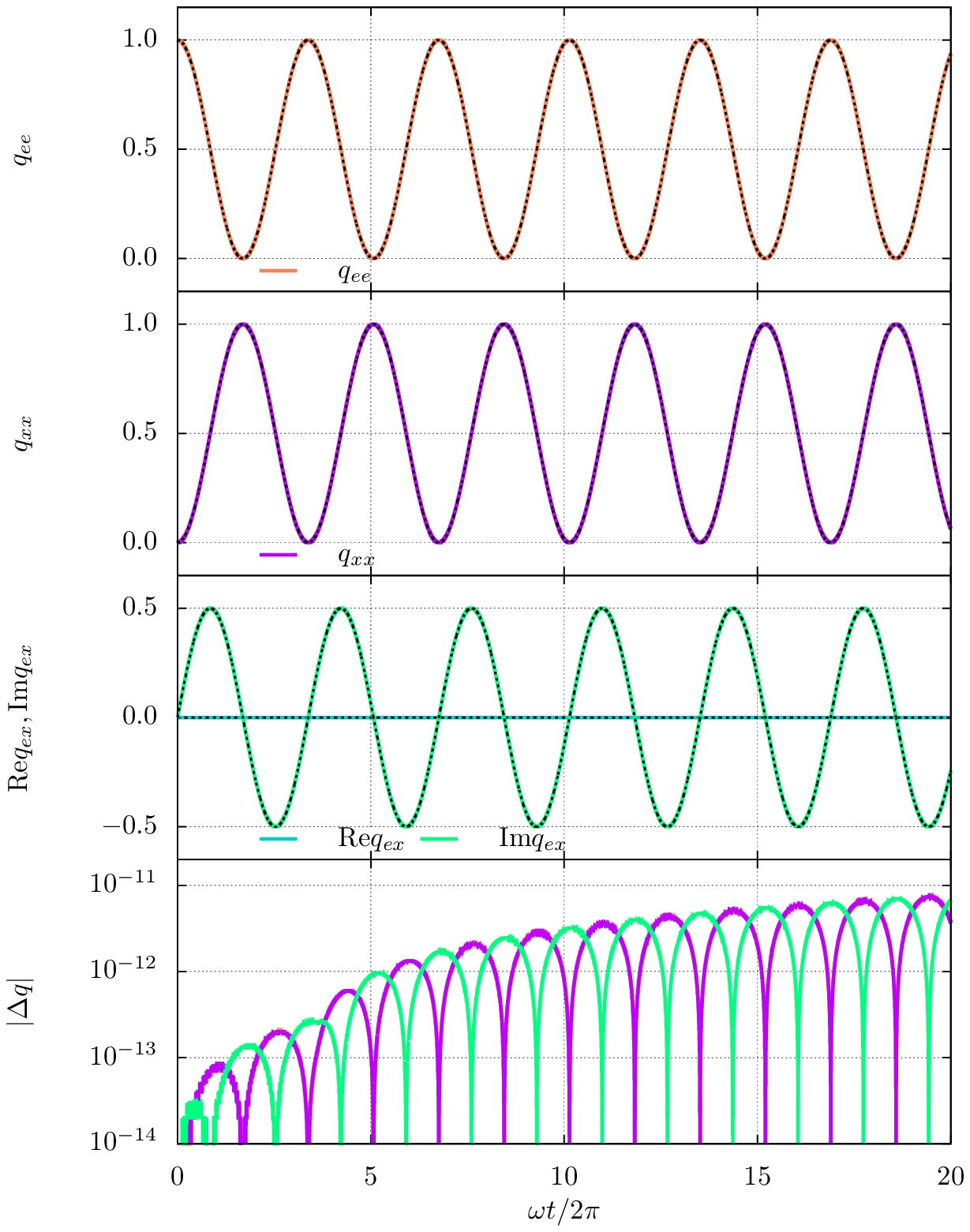}
    \includegraphics[width=\columnwidth]{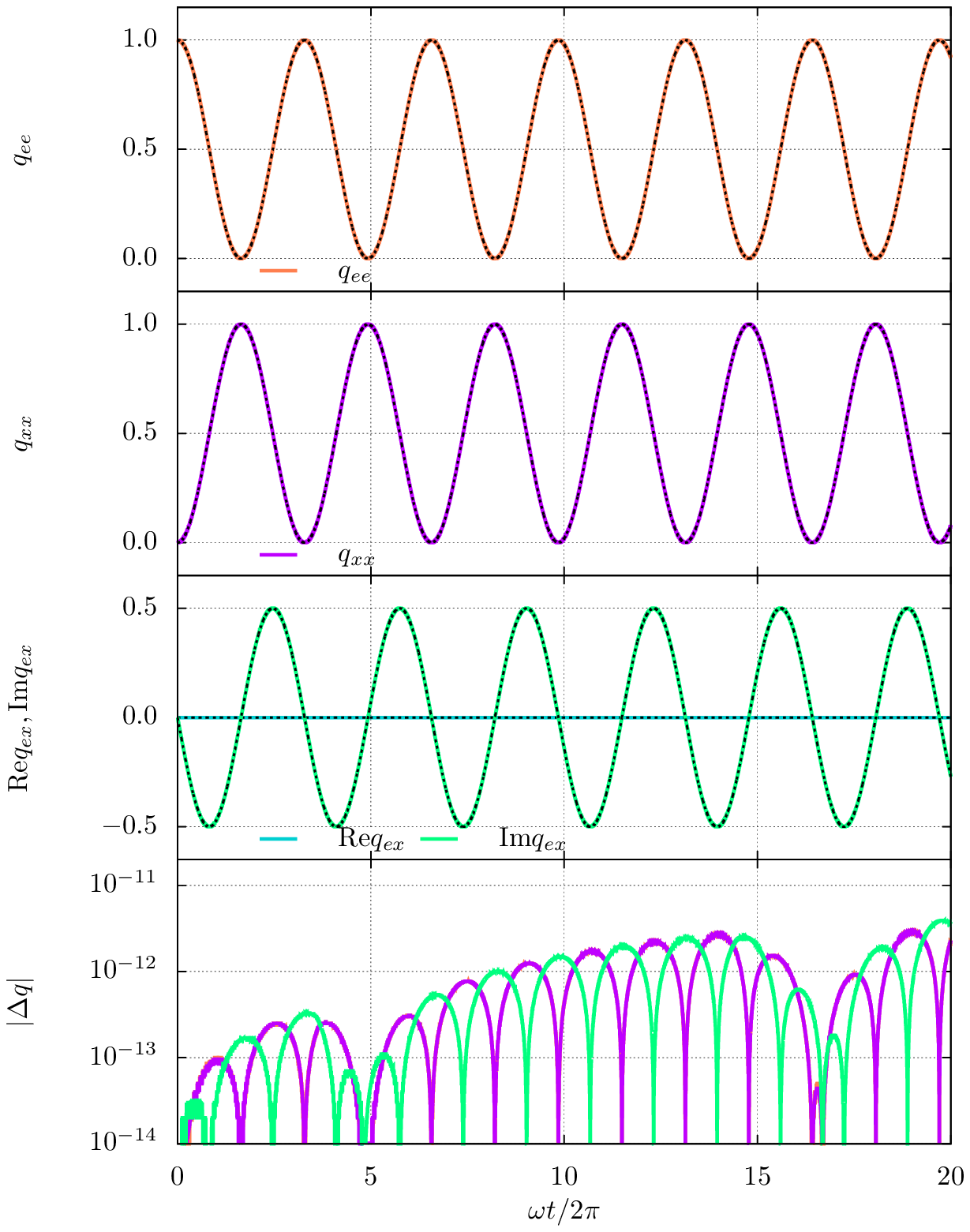}
    \caption{The same as Fig.~\ref{vac_NH} but for MSW effects. We adopt $n_e = n_{e0}$ (resonant oscillation). Left and right panels show the results with normal and inverted mass ordering, respectively.}
    \label{MSW_NH}
\end{figure*}

\begin{figure}[htbp]
    \centering
    \includegraphics[width = \columnwidth]{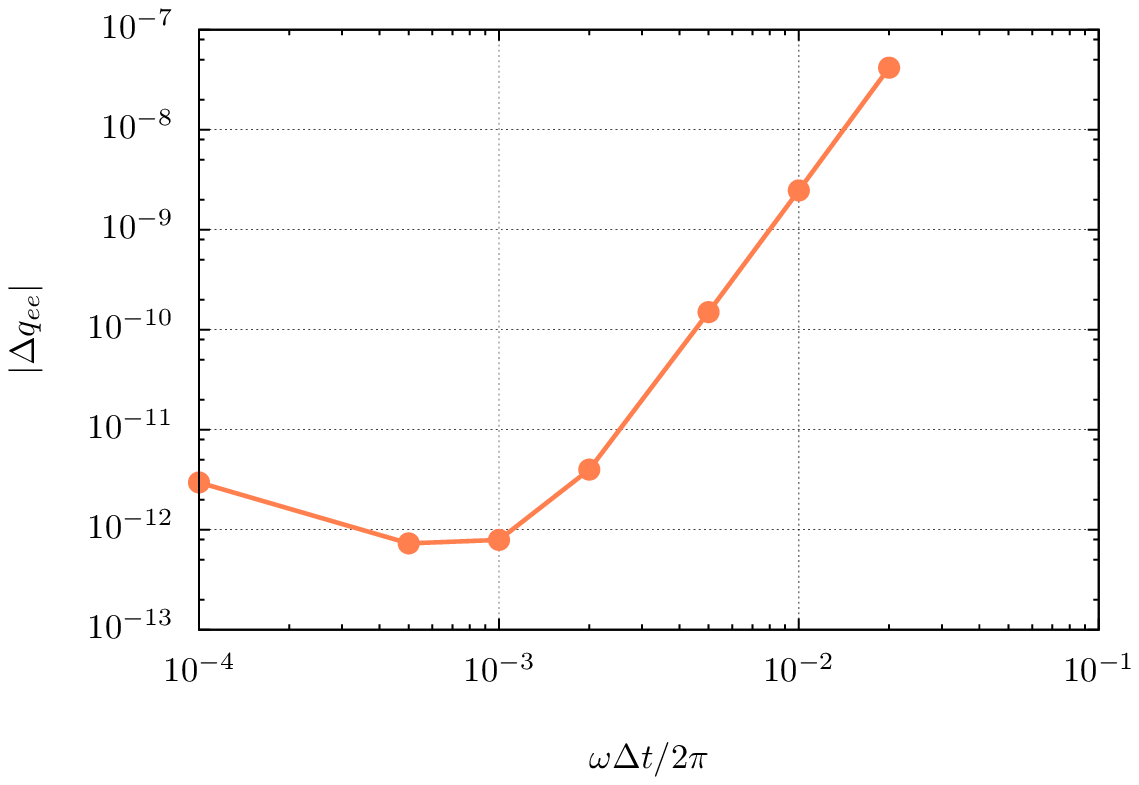}
    \caption{The same as Fig.~\ref{vac_error} but for a MSW test. We adopt $n_e = n_{e0}$ (resonant oscillation) in these simulations.}
    \label{MSW_error}
\end{figure}

It may be interest to see the resonance case ($n_e = n_{e0}$) in more detail. Figs.~\ref{MSW_NH} display the same quantities as displayed in Figs.~\ref{vac_NH}.
Fig.~\ref{MSW_error} shows the result of convergence check.
The error is measured at $t=2\pi/\omega \times10$, which is the same as that in the vacuum test (see also Fig.~\ref{vac_error}). These tests suggest that the MSW resonance can be handled accurately in our QKE-MC solver.

\section{Collective oscillation} \label{ch6}
We perform three representative tests for collective neutrino oscillations.
One of them is a collective neutrino oscillation in isotropic angular distributions. This is the same problem as done in \cite{hannestad2006}. The result is summarized in Sec.~\ref{subsec:colletestiso}.
In Sec.~\ref{subsec:twobeam}, we test a two-beam collective oscillation following \cite{chakraborty2016}. We compare the results with the linear analysis quantitatively.
In Sec.~\ref{ch6-3}, we carry out a test of fast flavor conversion. We evolve the system, in which the anisotropic neutrino distribution with a single ELN-crossing is set as an initial condition, by using QKE-MC solver. 
It should be mentioned that, although the physical setup is exactly the same as one of the simulations in \citet{shalgar2021b}, we do not reproduce their results. To address the issue, another simulation is carried out for the same problem but using our independent code (finite-difference method), that strengthened the reliability of QKE-MC solver. We note that, thanks to the cooperation by the authors of \citet{shalgar2021b}, we identify the source of discrepancy. We finally reached an agreement with them.

\subsection{isotropic distribution}\label{subsec:colletestiso}
In this test, we assume that neutrinos and anti-neutrinos have the same isotropic distributions. Regarding the matter potential, we assume $\bar{n}_e=n_x = \bar{n}_x = 0$.
Following the convention of \cite{hannestad2006}, we describe the QKE by using polarization vectors (see also Eq.~\ref{eq:polarization}),
\begin{eqnarray}
 &&\frac{\partial}{\partial t} \bold{P^\prime} = \left[\omega \bold{B} + \lambda \bold{L} + \mu \left(\bold{P^\prime} - \bar{\bold{P}}^\prime \right) \right] \times \bold{P^\prime}, \nonumber \\
 &&\frac{\partial}{\partial t} \bold{\bar{P}}^\prime = \left[-\omega \bold{B} + \lambda \bold{L} + \mu \left(\bold{P^\prime} - \bar{\bold{P}}^\prime \right) \right] \times \bold{\bar{P}}^\prime,
 \label{eq:poralization_iso}
\end{eqnarray}
with  $\lambda = \sqrt{2}G_F n_e$, $\mu = \sqrt{2}G_F n_\nu$ ($n_{\nu} = n_{\nu_e} = n_{\bar{\nu}_e}$) and  $\bold{L} = {\rm diag}[1,0]$. In the expression, we normalize the polarization vectors as $\bold{P}^\prime=\bold{P}/n_\nu$ and $\bold{\bar{P}}^\prime=\bold{\bar{P}}/n_\nu$.

As the first test, we ignore the matter potential, i.e., $\lambda$ is set to be zero.
In this case, Eq.~\ref{eq:poralization_iso} can be rewritten as,
\begin{eqnarray}
\bold{\dot{Q}} &=& \mu\bold{D} \times \bold{Q}, \nonumber \\
\bold{\dot{D}} &=& \omega \bold{B} \times \bold{Q},
\end{eqnarray}
with $\bold{Q} = \bold{P} + \bold{\bar{P}} - \omega \bold{B}/\mu$ and $\bold{D} = \bold{P}-\bold{\bar{P}}$.
As shown in \citet{hannestad2006}, we have four conserved quantities in the system: $|\bold{Q}|$, $\bold{D}\cdot\bold{B}$, $\bold{D}\cdot\bold{Q}$, $H\equiv\omega \bold{B}\cdot{Q} + \mu|\bold{D}|^2/2$. They are useful to check the accuracy of our QKE-MC solver.
Setting the initial condition with $\bold{P}^\prime(0)=\bold{\bar{P}^\prime}(0)=(0,0,1)$, we obtain
\begin{eqnarray}
\bold{D} = \frac{1}{\mu} \frac{\bold{Q} \times \bold{\dot{Q}}} {|\bold{Q}|^2}. \label{eq:Dofisocheck}
\end{eqnarray}
Eq.~\ref{eq:Dofisocheck} is an analogous equation to that describing the dynamics of charged particles propagating in electric field. It can be further cast into the following form,
\begin{eqnarray}
\ddot{\phi} = -\omega \mu Q \sin{\left(\phi + 2\theta_0\right)}, \label{eq:pendulum}
\end{eqnarray}
with $\bold{Q} = Q(\sin{\phi},0,\cos{\phi})$. This suggests
that the dynamics is analogous to a pendulum. We refer readers to \cite{hannestad2006} for more detail.

In the numerical setup of QKE-MC solver, we inject one MC particle with $E_\nu=20$ MeV initially with purely $\nu_e$ and $\bar{\nu}_e$ states.
We note that, their flight direction can be arbitrary chosen, meanwhile we ignore the angular dependence of neutrinos in the evaluation of the neutrino self-interaction potential during the time evolution.
By virtue of the prescription, only the isotropic component of the potential remains. 
It should be noted that it is possible to set isotropic angular distributions of neutrinos by generating many MC particles,
and compute the self-interaction potential without the above prescription. On the other hand, neutrino flight directions of each MC particle are constant with time, and the neutrino oscillation is identical among all MC particles. This indicates that computations of many MC particles are wasteful. Our prescription does not compromise the analysis and saves the computational resource. The number of neutrinos, i.e., the $ee$ component of $q$ (and $\bar{q}$) is chosen so that $\mu$ becomes $10 \omega$ ($\Delta m^2 = -2.45\times10^{-15}\ {\rm MeV^2}$). The mixing angle is set as $\theta_0=0.01$ to reproduce the result of \citet{hannestad2006}. In this test, we adopt a constant time step with $\mu \Delta t = 10^{-3}$.

\begin{figure}[htbp]
    \centering
    \includegraphics[width=\columnwidth]{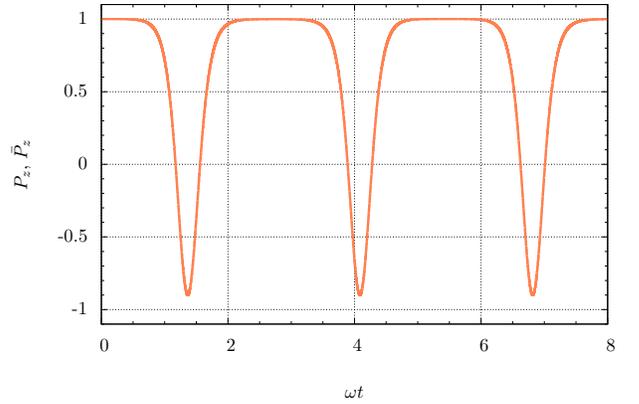}
    \caption{The time evolution of the z-components in polarization vectors in a test of isotropic collective neutrino oscillation without matter potential. We show the results for $\mu \Delta t=10^{-3}$ in this figure. See also Fig.1 in \cite{hannestad2006} that corresponds to a counterpart of this figure.}
    \label{hannestad_fig1}
\end{figure}

\begin{figure}[htbp]
    \centering
    \includegraphics[width=\columnwidth]{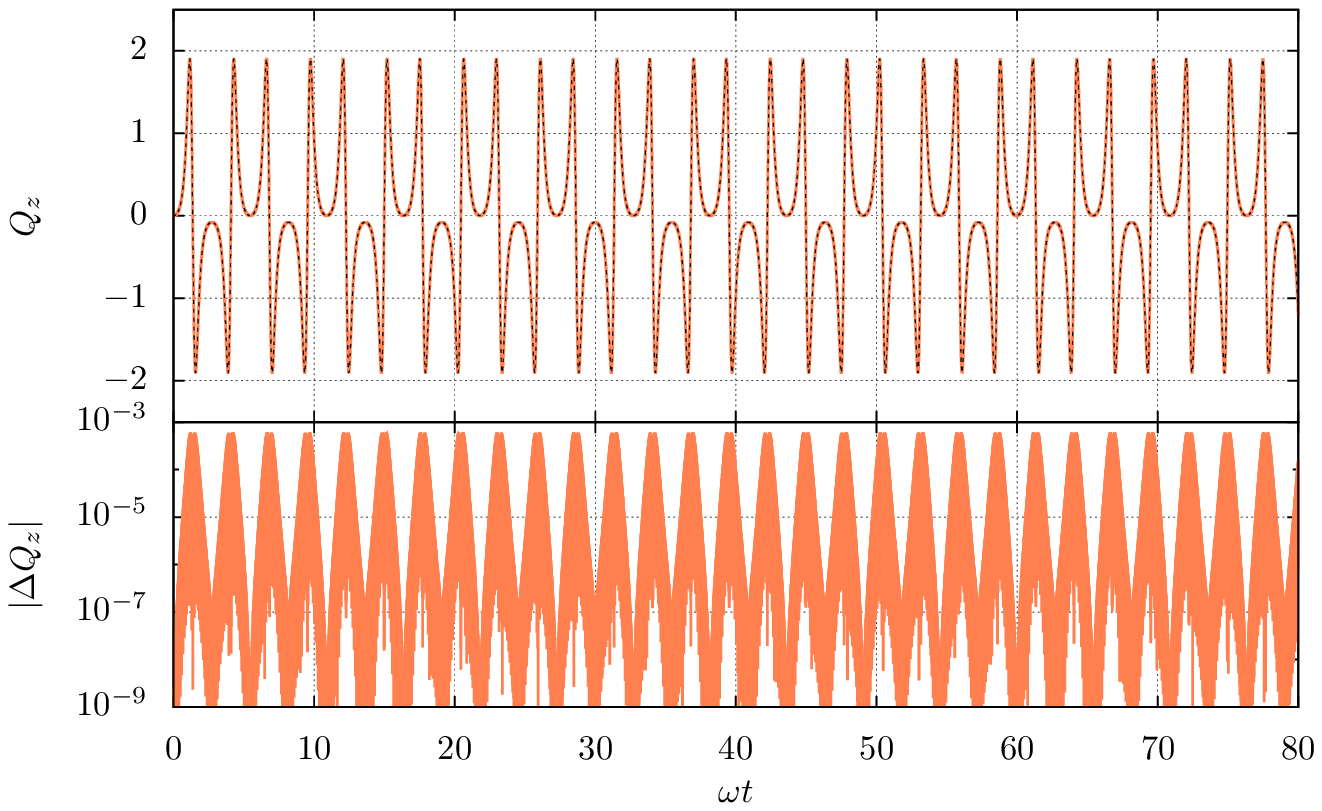}
    \caption{Time evolution of $Q_z$ for a test of isotropic collective neutrino oscillation without matter potential. In the top panel, we compare the results between QKE-MC (orange-solid line) and those obtained by solving Eq.~\ref{eq:pendulum} (black-dashed line).
    The difference, $|\Delta Q_z|$, is displayed in the bottom panel.
    We adopt a constant time step ($\mu \Delta t=10^{-3}$) in our QKE-MC simulation.}
    \label{hannestad_long}
\end{figure}

Figure~\ref{hannestad_fig1} portrays the time evolution of the z-components for $\bold{P}$ and $\bold{\bar{P}}$ computed by our QKE-MC solver. This is in good agreement with the result of \citet{hannestad2006} (see Fig.1 in their paper).
To see the capability of our code more quantitatively, we compare to the solutions obtained by solving Eq.~\ref{eq:pendulum}; the results are summarized in Fig.~\ref{hannestad_long}. We confirm that our QKE-MC solver provides a consistent result with the solutions of Eq.~\ref{eq:pendulum}; indeed, the deviation ($|\Delta Q_z|$) is $\lesssim 10^{-3}$ up to the time of $\omega t = 80$ (see the bottom panel of Fig.~\ref{hannestad_long}).

\begin{figure}[htbp]
    \centering
    \includegraphics[width=\columnwidth]{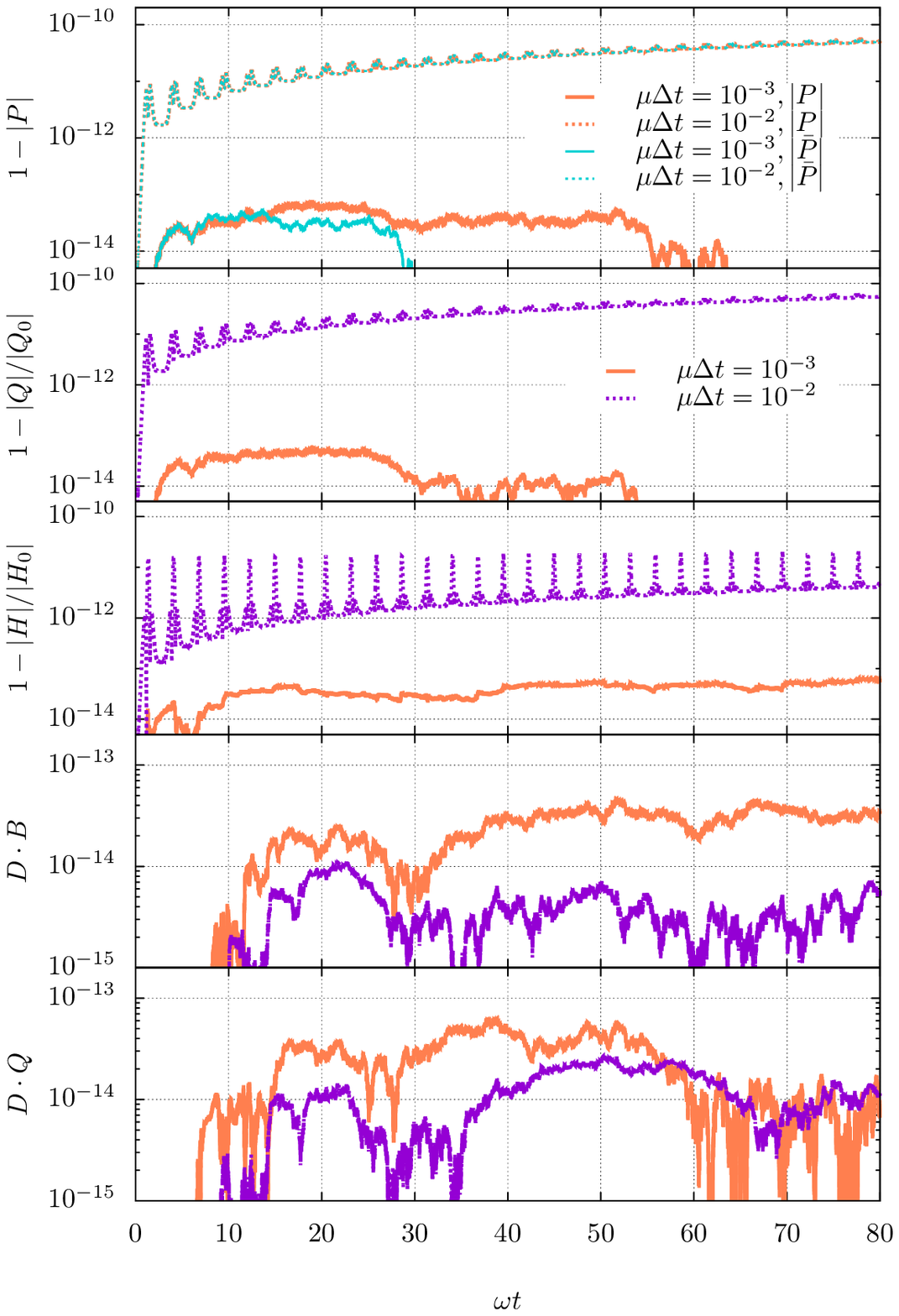}
    \caption{The time evolution of errors of conserved quantities for a test of isotropic collective neutrino oscillation without matter potential. From top to bottom, we display the results of $1-|\bold{P}|$, $1-|\bold{Q}|/|\bold{Q_0}|$, $1-H/H_0$, $\bold{D}\cdot\bold{B}$ and $\bold{D}\cdot\bold{Q}$,  respectively.
    We show the results with two different time steps in our QKE-MC simulation: $\mu \Delta t=10^{-3}$ (solid-line) and $10^{-2}$ (dashed-line).}
    \label{conserve_hannestad}
\end{figure}

Figure.~\ref{conserve_hannestad} displays the time evolution of conserved quantities. In this figure, we add the result with $\mu \Delta t = 10^{-2}$ to see the dependence of time step. All the quantities, or $|\bold{P}|$, $|\bold{Q}|$, $H$, $\bold{D}\cdot\bold{B}$ and $\bold{D}\cdot\bold{Q}$, are well conserved in our results, and we confirm that the smaller $\Delta t$ tends to improve the accuracy of the conserved quantities.

\begin{figure}[htbp]
    \centering
    \includegraphics[width=\columnwidth]{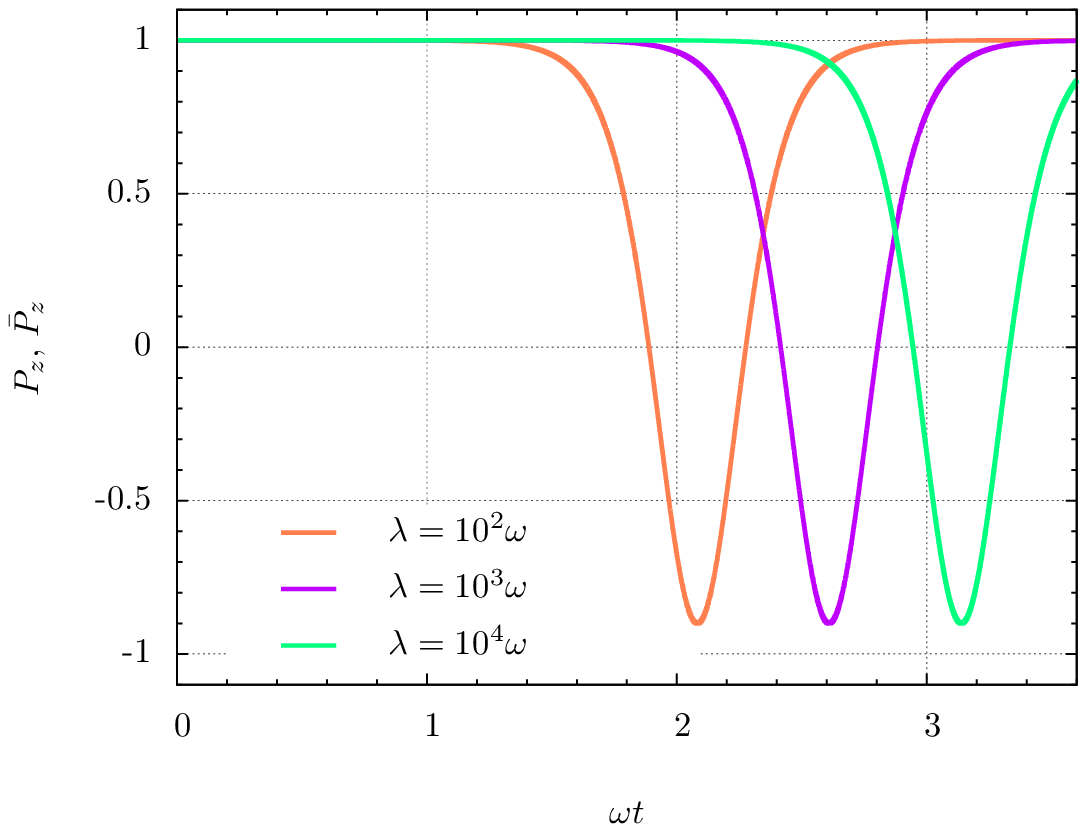}
    \caption{The same as Fig.~\ref{hannestad_fig1} but for the test including matter potential.
    Different colors denote results for different values of matter potential: $\lambda = 10^2 \omega, 10^3 \omega$, and $10^{4} \omega$. We adopt a constant time step ($\mu \Delta t=10^{-3}$) in this test .}
    \label{hannestad_mat}
\end{figure}

As another test of isotropic collective neutrino oscillations, we include the matter term in addition to the self-interaction potential (see Eq.~\ref{eq:poralization_iso}). Same as \citet{hannestad2006}, we vary the value of $\lambda$: $\lambda=10^2\omega$, $10^3\omega$ and $10^4\omega$.
Figure~\ref{hannestad_mat} shows the time evolution of $P_z$ and $\bar{P}_z$ obtained by our QKE-MC (see also Fig.~\ref{hannestad_fig1} for comparison to the case without matter potential).
We confirm that our results reproduce the same results reported in \cite{hannestad2006} (see Fig.2 in the paper). This suggests that our code is capable of handling the collective neutrino oscillations in isotropic distributions.

\subsection{Two-beam collective oscillation}\label{subsec:twobeam}

\begin{figure*}[htbp]
    \centering
    \includegraphics[width=\columnwidth]{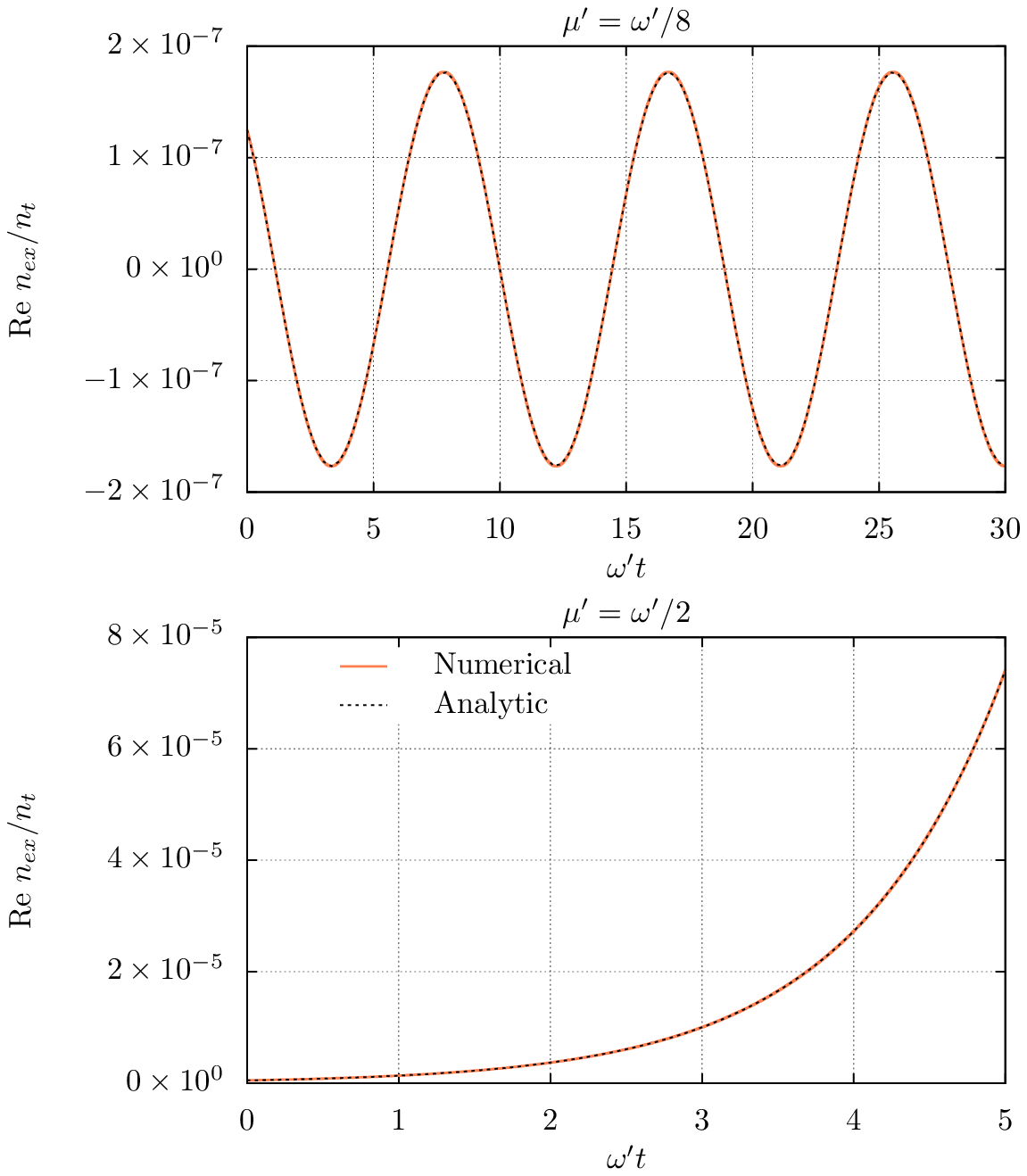}
    \includegraphics[width=\columnwidth]{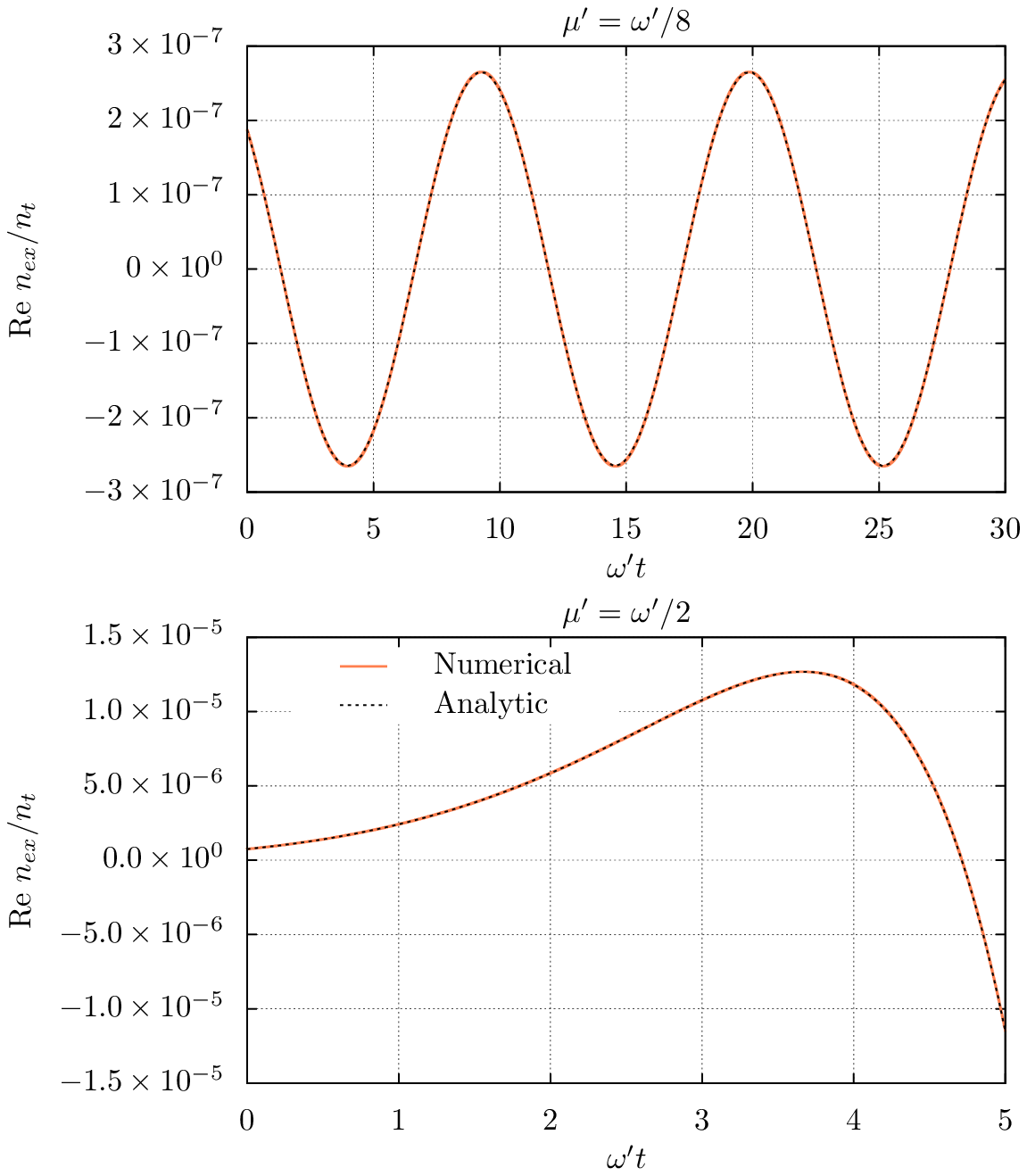}
    \caption{The time evolution of the real part of $n_{ex}$ in the test of two-beam collective oscillation. In the left and right panels, we adopt the different asymmetric parameter, $\alpha=0$ (left) and $0.5$ (right). Orange and black lines denote the QKE-MC results and the those computed from eigen frequency, respectively. We show the results with two different $\mu^\prime$'s in top and bottom panels. The top and bottom panels correspond to the stable and unstable system to flavor conversion. See text for more detail.}
    \label{two-beam}
\end{figure*}

We employ a two-beam model as a test of collective oscillation. In this model, neutrinos and anti-neutrinos, propagating in the opposite direction, intersect each other. The characteristics of flavor conversion has been well studied \citep[see, e.g.,][]{chakraborty2016}. We also note that the dispersion relation of the instability can be written in a concise way (see below). Hence, we compare the result of QKE-MC simulation to the analytic solution in the linear regime.

Following \citet{chakraborty2016}, we give an overview of the derivation of eigen frequency.
In the system, we obtain the time evolution of the number densities from QKE as,
\begin{eqnarray}
i \frac{\partial n_{ex,1}}{\partial t} &=& \left(-\frac{\Delta m^2}{2E_\nu}-2\sqrt{2}G_F\bar{n}_{ee,2}\right)n_{ex,1} \nonumber \\
&& \ \ \ \ \ \ \ + 2\sqrt{2}G_F\bar{n}_{ex,2}^\ast n_{ee,1},\\
i \frac{\partial \bar{n}_{ex,2}}{\partial t} &=& \left(-\frac{\Delta m^2}{2E_\nu}-2\sqrt{2}G_Fn_{ee,1}\right)\bar{n}_{ex,2} \nonumber \\
&&  \ \ \ \ \ \ \ + 2\sqrt{2}G_Fn_{ex,1}^\ast \bar{n}_{ee,2}, \label{eq:twobeambase}
\end{eqnarray}
where indices 1 and 2 distinguishes the flight direction of neutrinos, i.e., specifying neutrinos and anti-neutrinos, respectively; we assume $\theta_0 = 0$ and a monochromatic neutrino energy with $E_\nu$. It should be mentioned that we integrate over the momentum space in these equations.
Before linearizing the equation, we define two new variables, $g$ and $S$, as
\begin{eqnarray}
n_i - \bar{n}_i^\ast = \frac{1}{2}{\rm Tr}\left(n_i-\bar{n}_i^\ast \right) 
    + \frac{g_i}{2}\left(
     \begin{array}{cc}
      1 & S_i\\
      S_i^\ast & -1
    \end{array} \right),
\end{eqnarray}
where $g_i = n_{ee,i}-n_{xx,i}-\bar{n}_{ee,i}+\bar{n}_{xx,i}$.
In our test case, $g_1 = n_{ee,1}$, $g_2 = -\bar{n}_{ee,2}$, $n_{ex,1}=g_1S_1/2$ and $\bar{n}_{ex,2} = -g_2S_2^\ast/2$.
By using the new variables, the linearized equations can be written as
\begin{eqnarray}
i \frac{\partial S_1}{\partial t} &=& \left(-\frac{\Delta m^2}{2E_\nu}+2\sqrt{2}G_Fg_2\right)S_1 - 2\sqrt{2}G_Fg_2S_2, \\
i \frac{\partial S_2^\ast}{\partial t} &=& -\left(\frac{\Delta m^2}{2E_\nu}+2\sqrt{2}G_Fg_1\right)S_2^\ast + 2\sqrt{2}G_Fg_1S_1^\ast .\ \ \ \ \ \ \ 
\end{eqnarray}
Adopting plane wave expansion with $S_i = Q_i\exp{(-i\Omega t)}$, the eigen frequency can be written as,
\begin{eqnarray}
\Omega &=& \sqrt{2}G_F\left(g_1 + g_2\right) \nonumber \\
 && \pm \sqrt{2G_F^2\left(g_1 + g_2 \right)^2 + \omega^\prime \left(\omega^\prime + 2\sqrt{2}G_F\left(g_1-g_2\right) \right)}, \ \ \ \ \ \ \ \ 
\end{eqnarray}
with $\omega^\prime = -\omega$.
In this test, neutrino distributions are assumed to be homogeneous \citep[see][for more general cases]{chakraborty2016}.
For convenience, we introduce an asymmetry parameter between neutrinos and anti-neutrinos, $\alpha$. We rewrite the number densities of neutrinos and anti-neutrinos by using $\alpha$ as $n_{ee}=(1+\alpha)n_t$ and $\bar{n}_{ee}=(1-\alpha)n_t$, respectively, where $n_t$ represents their average, i.e., $n_t = 0.5(n_{ee}+\bar{n}_{ee})$.
The dispersion relation can be rewritten as,
\begin{eqnarray}
\Omega_{\pm} = 2 \alpha \mu^\prime \pm \sqrt{\left(2 \alpha \mu^\prime \right)^2 + \omega^\prime\left(\omega^\prime-4\mu^\prime \right)},
\end{eqnarray}
with $\mu^\prime = \sqrt{2}G_F n_t$.
The imaginary part of $\Omega$ has a non-zero value in cases with $\left(2 \alpha \mu^\prime \right)^2 + \omega^\prime \left(\omega^\prime -4\mu^\prime \right) < 0 $. In this condition, the slow flavor instability kicks in.
In other cases, the flavor conversion occurs periodically with constant frequencies.
The corresponding eigenvector of $\Omega_{-}$, for example, can be written as
\begin{eqnarray}
 \left( \begin{array}{c}
  Q_1 \\ Q_2
 \end{array} \right)
 = 
 \left( \begin{array}{c}
  2\mu^\prime \left(\alpha -1\right) \\ \omega^\prime +2\mu^\prime \left(\alpha -1\right)-\Omega_{-}
 \end{array} \right) b. \label{eigenvecOmegamin}
\end{eqnarray}
This is used for an initial condition in our QKE-MC solver.

To reproduce the two-beam environment in our QKE-MC simulation, we prepare two MC particles. One of them is for neutrinos, and the other is for anti-neutrinos. The flight direction of each particle is opposite each other. The neutrino energy is set to be $E_\nu = 50$ MeV. To determine $\Delta t$, we compare the vacuum- and collective oscillation timescale at each time step, and then adopt the smaller one. We then multiply it by a factor of $10^{-4}$. We demonstrate four simulations by changing $\alpha$ ($\alpha=0$ and $0.5$) and $\mu^{\prime}$ ($\mu^\prime=\omega^\prime/8$ and $\omega^\prime/2$). In this test, we adopt the value of mass difference: $\Delta m^2 = -2.45\times10^{-3}{\rm eV}^2$. The corresponding time scale of $1/\omega$ is 2.69$\times 10^{-5}$ s, and the number density of $n_t$ with $\mu^{\prime} = \omega^\prime$ is $1.93\times10^{26}{\rm cm^{-3}}$. The initial condition of $q$ and $\bar{q}$ is set so as to coincide with an eigenvector with $\Omega_{-}$ with ${\rm Re}b = {\rm Im}b = 10^{-6}/2\mu^\prime(\alpha-1)$ in Eq.~\ref{eigenvecOmegamin}.

Figure~\ref{two-beam} shows the time evolution of the real part of $n_{ex}$ in the linear regime ($|S| \ll 1$).
Orange and black lines represent the QKE-MC results and those computed from the linearlized QKE.
Top and bottom panels show the results for $\mu^\prime=\omega^\prime/8$ and $\omega^\prime/2$, where the former and latter systems are the stable and unstable with respect to slow flavor conversion.
Regardless of the asymmetric parameter and stability, 
our QKE-MC code can accurately reproduce the analytic solutions in the linear phase.


\subsection{fast flavor conversion in axisymmetric momentum space} \label{ch6-3}

As a test of fast flavor conversion, we carry out simulations with the same physical condition of \citet{shalgar2021b}, in which the angular distributions of $\nu_e$ and $\bar{\nu}_e$ has initially a single crossing in momentum space, indicating that it is unstable to fast flavor conversion. We apply our QKE-MC solver to evolve the system under homogeneous approximation, and then we compare the results to those in \citet{shalgar2021b}.

In this test, we assume a monochromatic neutrino energy with $E_{\nu}$ and axial symmetry in momentum space. For convenience, we define the azimuthal-integrated density matrix as,
\begin{eqnarray}
\rho_a(\theta_\nu)= \int \rho(\theta_\nu,\phi_\nu) d\phi_\nu. \label{eq:azimueneinterho}
\end{eqnarray}
The governing equation in the system can be written as
\begin{eqnarray}
i\frac{\partial \rho_a}{\partial t} &=& [H_{\rm vac}^\prime+H_{\rm mat} + H_{\nu\nu}^\prime,\rho_a], \nonumber \\
i\frac{\partial \bar{\rho}_a}{\partial t} &=& [H_{\rm vac}^{\prime}-H_{\rm mat}^\ast - H_{\nu\nu}^{\prime\ast},\bar{\rho}_a],
\end{eqnarray}
with 
\begin{eqnarray}
H_{\rm vac}^\prime &=& U\frac{1}{2E_{\nu}}
\left(
    \begin{array}{cc}
      m_1^2 & 0\\
      0 & m_2^2 
    \end{array}
  \right)
U\dagger, \nonumber \\
H_{\nu\nu}^\prime &=& \sqrt{2}G_F  \int \frac{d\cos{\theta_\nu^\prime}}{\left(2\pi \right)^3} \left(1-\cos{\theta_\nu}\cos{\theta_\nu^\prime}\right) \nonumber \\  && \times \left(\rho_a^\prime-\bar{\rho}_a^{\prime\ast}\right). 
\end{eqnarray}
We set an initial condition,
\begin{eqnarray}
\rho_{ee,a0} &=& 0.5n_\nu, \nonumber \\
\bar{\rho}_{ee,a0} &=&  \left[ 0.47+0.05\exp{\left(-\left(\cos{\theta_\nu}-1\right)^2\right)} \right] n_\nu, \label{eq:inifast}
\end{eqnarray}
where $n_{\nu}$ denotes the total number density of $\nu_e$. We assume $\mu=10^5{\rm km^{-1}}$, $\theta_0 = 10^{-6}$, $\Delta m^2 = 2.50\times 10^{-6}{\rm eV^2}$, and $E_{\nu}=50$ MeV, which leads to $\omega = 1.27 \times10^{-4} {\rm km}^{-1}$. We note that this setup is the same as the case A in \citet{shalgar2021b}.

For convenience, we provide the time scale under the choice of these parameters. They are $3.33\times10^{-11}$s and $2.67\times10^{-2}$s for neutrino self-interaction and vacuum oscillation, respectively. Since $\mu \gg \omega$ is satisfied, the neutrino oscillation is dominated by fast flavor conversion\footnote{However, the vacuum term is important to generate a seed perturbation in the system. Since there are no neutrinos in the mixing flavor-states at initial time, the flavor conversion does not occur without the vacuum contribution.}.

In QKE-MC simulations, we inject MC particles isotropically. To see the convergence, we change the total number of MC particle as 64, 128, 256 and 512. Following Eq.~\ref{eq:inifast}, we determine the values of $q$ and $\bar{q}$ in each MC particle. During the time evolution, $\Delta t$ is not constant. It is essentially determined by the maximum difference of $\rho_a$ and $\bar{\rho}_a$ at each time step. We compute the effective ELN number density by assuming that the maximum difference is distributed isotropically. This gives a conservative time scale of the fast flavor conversion. We further divide it by a factor of $dt_{\rm fac}$. We adopt $dt_{\rm fac} = 50$ as a representative case. It should be mentioned that the MC particles do not change their momentum state in this test due to no collision terms, indicating that the statistical noise is not an issue in this problem. Hence, we do not need a special treatment to reduce the noise (see also Sec.~\ref{subsec:col}).

\begin{figure}[htbp]
    \centering
    \includegraphics[width=\columnwidth]{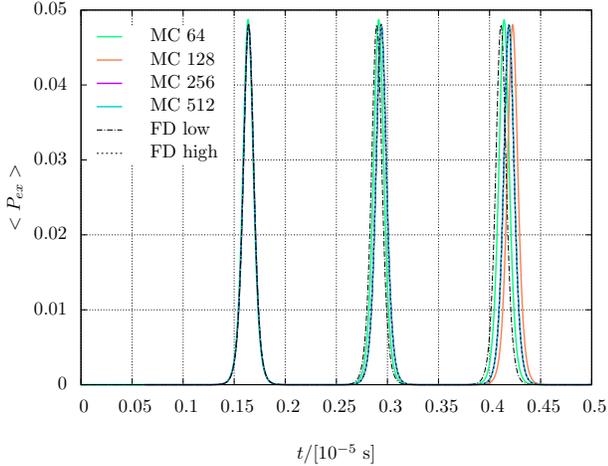}
    \caption{The time evolution of transition probability $<P_{ex}>$ for a test of fast flavor conversion (Sec.~\ref{ch6-3}). Solid lines represent QKE-MC results and the color distinguishes the total number of MC particles. Black lines denote results of finite-difference solver. The line type distinguishes the resolution. The high resolution case adopts 256 angular grids and $\Delta t=0.5{\rm cm}/c$, while the low resolution one uses 128 angular grids and $\Delta t=1{\rm cm}/c$. $c$ denotes the speed of light in cgs unit.}
    \label{morinaga}
\end{figure}

Figure~\ref{morinaga} shows the time evolution of a transition probability, which is defined by
\begin{eqnarray}
<P_{ex}> = 1-\frac{\int \rho_{ee,a} d\cos{\theta_\nu}}{\int \rho_{ee,a0} d\cos{\theta_\nu}}.
\end{eqnarray}
Solid lines represent our results and color distinguishes the total number of MC particles.
In the linear phase ($\lesssim 10^{-6}$ s), the resolution dependence (with respect to the total number of neutrinos) is minor; indeed the first peak profile in Fig.~\ref{morinaga} is almost identical among all simulations. However, the deviation is obvious at the latter phase; the peak profile in the low resolution tends to appear earlier. On the other hand, we find no clear differences between the simulation of 256 and 512 MC particles up to the third peak.

In \citet{shalgar2021b}, there is a counterpart figure, the blue line in the top left panel of Fig.~2 in their paper, to our Fig.~\ref{morinaga}. By comparing between the two results, the flavor conversion in their simulation tends to evolve slower than ours; for instance,
the first peak of $<P_{ex}>$ appears at $t\sim 0.6 \times 10^{-5}$ s. 
We address the issue by the following way. At first, we perform another simulation, in which we solve QKE with poralization vectors (see Eq.~\ref{eq:polarization}) by using finite-difference method. This is a deterministic- and mesh-based solver, indicating that the numerical artifacts inherent in our MC solver are not involved. 
We carry out two simulations with two different resolutions; the high resolution simulation adopts $\Delta t= 0.5{\rm cm}/c$ and 256 angular grids, while the low resolution one adopts $\Delta t=1{\rm cm}/c$ and 128 angular grids.
If the MC code has some issues, the detailed comparison between the two independent code illustrates the origin of the discrepancy from \citet{shalgar2021b}.

We found, however, that the finite difference method shows a consistent result with QKE-MC, which can be seen in Fig.~\ref{morinaga} (black dashed- and dashed lines).
It should also be mentioned that the difference between both codes becomes smaller with increasing resolutions that lends confidence to our QKE-MC code. In the wake of the result, we have contacted the authors of \citet{shalgar2021b}, and they mentioned that there are typos in the paper\footnote{In their simulation, the neutrino self-interaction potential was four time smaller than that described in \citet{shalgar2021b} (private communication).}, and they reproduced our results. Hence, the issue of discrepancy is completely resolved.

\begin{figure}[htbp]
    \centering
    \hspace*{-1.0cm}
    \includegraphics[width=9.5cm]{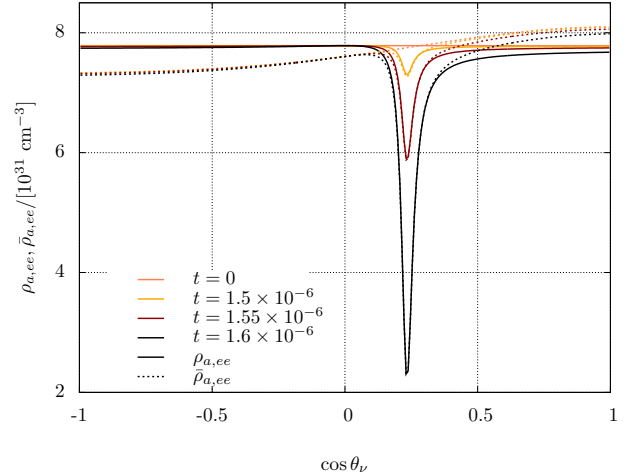}
    \caption{The time evolution of neutrino angular distributions for a test of fast flavor conversion (Sec.~\ref{ch6-3}). We display the result up to the time of the first peak. Color distinguishes the time. Solid and dashed lines denote $\rho_{a,ee}$ and $\bar{\rho}_{a,ee}$, respectively.
    In this figure, we show the result of QKE-MC simulation with 256 MC particles.}
    \label{angle_dist}
\end{figure}

To see the flavor conversion in more detail, we display the time evolution of neutrino angular distributions computed by our QKE-MC solver up to the time of the first peak in Fig.~\ref{angle_dist}.
In this figure, we show the result with 256 MC particles.
As shown in the plot, the flavor conversion occurs around the ELN crossing point; the qualitative trend is the same as that found in \citet{shalgar2021b}.

\begin{figure}[htbp]
    \includegraphics[width=\columnwidth]{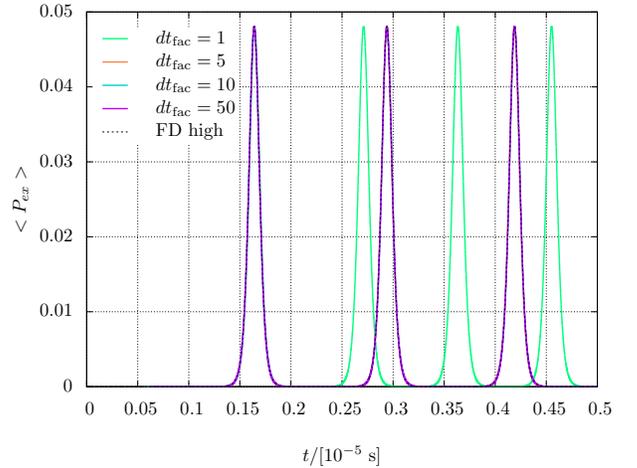}
    \caption{The $\Delta t$ dependence of $<P_{ex}>$ evolution.
    Different colors denote the results with different time resolution: $dt_{\rm fac} = 1, 5, 10$ and 50 (see text for the definition of $dt_{\rm fac}$). For comparison, the result of the high resolution simulation of finite-difference solver is also shown in black dashed line.}
    \label{fast_dt_depend}
\end{figure}

Finally, we check the dependence of $\Delta t$, that can be controlled by $dt_{\rm fac}$. The results are displayed in Figure~\ref{fast_dt_depend}. We vary $dt_{\rm fac}$ from 1 to 50. We find that $dt_{\rm fac}=5$ provides the sufficient time resolution to this model. This also provides the consistent result with that obtained by the high resolution simulation of finite-difference method.

\section{QKE with scattering process} \label{ch7}

In this section, we apply our QKE-MC solver to problems including collision term. We pay special attention to scattering processes in this paper; there are mainly two reasons. One of them is that the scattering process is essentially the combined process of emission and absorption, indicating that this test provides a good validation for these processes. The other reason is that roles of scattering in collective neutrino oscillations is gaining increased attention from the community \citep{shalgar2021b,martin2021,johns2021}. Following the previous studies, we focus on an isoenergetic scattering in QKE. We note that the blocking factor in isoenergetic scatterings is exactly canceled between in- and out-scatterings\footnote{The blocking factor needs to be evaluated appropriately when we apply our code to emission-, absorption- and non-isoenergetic scattering processes. To do this, the energy volume in momentum space needs to be taken into account (to compute the density matrix), indicating that the monochromatic approximation can not be applied in the system. In fact, the delta function in the energy spectrum of neutrinos (corresponding to the monochromatic approximation) violates the characteristics of Fermions.}. Although we postpone more detailed investigations on QKE with collision terms, this study has a role of the pilot study for them.

\subsection{Test for scattering module} \label{ch7-2}
Before presenting a code test for a coupling system between neutrino oscillations and scatterings, we carry out another test to see the capability of scattering module implemented in our QKE-MC solver. As we described in Sec.~\ref{subsec:col}, we change numerical treatments of scattering from our classical MC solver, indicating that the capability needs to be tested. We perform a cross-validation study by using our classical MC solver.

The detail of this test is as follows. We set anisotropic neutrino distributions on the monochromatic neutrino energy as an initial condition. The angular distribution of neutrinos is the same as that used for $\bar{\nu}_e$ in another test of fast flavor conversion (see Eq.~\ref{eq:inifast} in Section~\ref{ch6-3}). We follow the evolution of the system by using two codes: QKE-MC solver and classical MC solver. We consider the homogeneous system and incorporate isoenergetic scattering (see below for the reaction rate). We note that the oscillation module is turned off in the simulation by the QKE-MC solver.

In QKE-MC solver, we adopt 2,000 MC particles. In each particle, $ee$ component of $q$ is finite but others are zero.
The inverse mean free path ($C$) is set as $0.05\ {\rm km^{-1}}$.
In this simulation, we apply EMFP method to reduce the statistical noise. We 
vary the parameter $a$ as $10^{-2}$, $10^{-3}$ and $10^{-4}$. We note that simulations with the smaller value of $a$ reduce more noise but become more computationally expensive (since the frequency of scattering for each MC particle increases with decreasing $a$, see Sec.~\ref{subsec:col} for the detail.). On the other hand, the memory capacity is not pressed, that enables us to run the simulation on standard workstations. To reduce the computational cost, the newly generated MC particles due to scatterings on each time step are combined promptly to the adjacent ones. Although this prescription smears out the small angular structure of neutrinos, such a detailed structure is out of the scope of this test.
In the classical MC transport \citep{kato2020}, on the other hand, 
we use $10^5$ MC particles and run 200 simulations with the identical initial condition. The resultant statistics is high enough to quantitatively capture the time evolution of angular distributions of neutrinos.

\begin{figure}[htbp]
    \centering
    \includegraphics[width=\columnwidth]{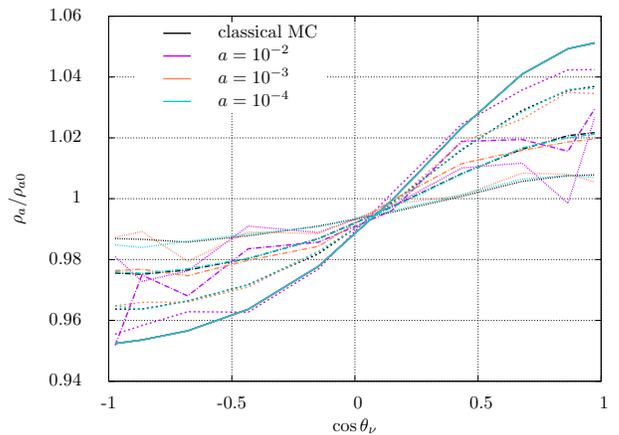}
    \caption{The time evolution of neutrino angular distributions in a test of scattering module (Sec.~\ref{ch7-2}). Color distinguishes the model: purple ($a=10^{-2}$ with QKE-MC), orange ($a=10^{-3}$ with QKE-MC), light blue ($a=10^{-4}$ with QKE-MC), and black (classical MC). Line-type distinguishes the time snapshot: $t/t_m$=0, 0.3, 0.75, 1.5, where $t_m$ denotes the mean free time ($t_m=6.67\times10^{-5}$ s).
    }
    \label{comp_scat}
\end{figure}

Figure~\ref{comp_scat} shows the time evolution of the neutrino angular distribution, illustrating that scatterings isotropize the angular distribution of neutrinos. We find that the statistics is improved with reducing $a$. The result with $a=10^{-4}$ is almost identical to that obtained from classical MC solver. This is a strong evidence that the EMFP method works well without degrading physical fidelity.

\subsection{Fast flavor conversions with scatterings} \label{ch7-3}

\begin{figure}[htbp]
    \centering
    \hspace*{-0.9cm}
    \includegraphics[width=9cm]{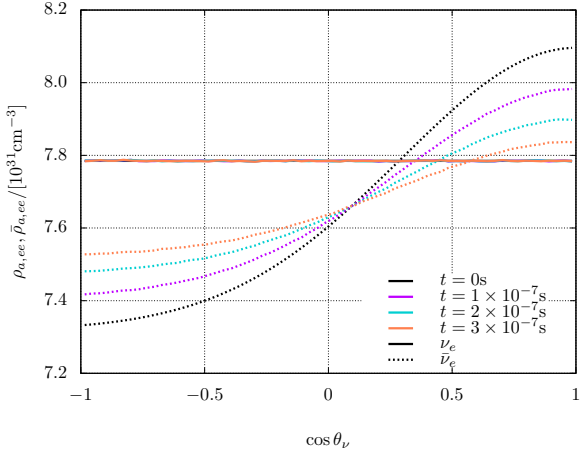}
    \caption{The time evolution of the angular distributions for $\nu_e$ (solid) and $\bar{\nu}_e$ (dashed) for a test of fast flavor conversion with scatterings. 
    Different colors denote different time steps.}
    \label{angle_e}
\end{figure}

As a final test, we demonstrate simulations for fast flavor conversion with scatterings.
As shown in Sec.~\ref{ch6-3}, however, our simulations did not reproduce the published result of \citet{shalgar2021b} in cases without scatterings\footnote{As mentioned in Sec.~\ref{ch6-3}, the inconsistency to the results of \citet{shalgar2021b} was resolved by making detailed comparison between the two results; see text for more detail.}.
We, hence, take another way to validate our results.
In this test, we focus on the early phase (the flavor conversion growth exponentially), and provide clear physical interpretations to the dynamics. For a quantitative assessment, we compare the growth rate of flavor conversion in our QKE-MC simulation to an empirical (analytic) formula given by \citet{morinaga2020}.

We start with describing the governing equation of the system. Following the convention used in Sec.~\ref{ch6-3}, we rewrite the QKE as,
\begin{eqnarray}
i\frac{\partial \rho_a}{\partial t} &=& \left[H^\prime_{\rm vac} + H_{\rm mat} + H^\prime_{\nu\nu}, \rho_a\right] \nonumber \\
&&+ i\int_{-1}^1 \tilde{C}\rho_a^\prime d\cos{\theta_\nu^\prime} \nonumber \\
&&- i\int_{-1}^1 \tilde{C}\rho_a d\cos{\theta_\nu^\prime}, \\
i\frac{\partial \bar{\rho}_a}{\partial t} &=& \left[H^\prime_{\rm vac} - H^\ast_{\rm mat} - H^{\prime\ast}_{\nu\nu}, \bar{\rho}_a\right] \nonumber \\
&&+ i\int_{-1}^1 \tilde{C}\bar{\rho}_a^\prime d\cos{\theta_\nu^\prime} \nonumber \\
&&- i\int_{-1}^1 \tilde{C}\bar{\rho}_a d\cos{\theta_\nu^\prime},\ \ \ \  
\end{eqnarray}
where $\tilde{C}$ represents a scattering rate.
In this test, we assume that $\tilde{C}$ is constant (no angular dependence). We note that the inverse mean free path $C$ is 2 times larger than $\tilde{C}$ in this expression \citep[see also][]{shalgar2021b} and we adopt  $C=10\ {\rm km}^{-1}$ in this study\footnote{This parameter corresponds to the highest rate among models in \citet{shalgar2021b}.}.
It should be mentioned that a large number of MC particles is necessary compared to the case without scatterings (see Sec.~\ref{ch6-3}). This is due to the fact that the statistical noise makes the simulation troublesome in this test. Following the prescriptions presented in Sec.~\ref{subsec:col}, we reduce the noise by a smoothing method (in computations of self-interaction potential) and EMFP method. As a reference, we adopt the number of MC particles, the number of angular grids ($N_\theta$) for the smoothing, and EMFP parameter ($a$) are 32,000, 64, and $10^{-5}$, respectively. We show that this provides a sufficient resolution up to the time of end of our simulation ($3\times10^{-7}$~s). The resolution dependence is also discussed in this section.

Figure~\ref{angle_e} shows the time evolution of the angular distributions for $\nu_e$ and $\bar{\nu}_e$.
The former has an isotropic distribution initially, whereas the latter has an anisotropic distribution (see black lines).
This figure illustrates that $\bar{\nu}_e$ angular distribution is isotropized with time, and the ELN crossing point changes with time (approaching to $\cos \theta_{\nu} = 1$). On the other hand, $\nu_e$ distribution does not evolve, indicating that the flavor conversion is subtle. This suggests that the remarkable change of angular distributions of $\bar{\nu}_e$ is due to scatterings.

\begin{figure}[htbp]
    \hspace*{-1.5cm}
    \includegraphics[width=9.8cm]{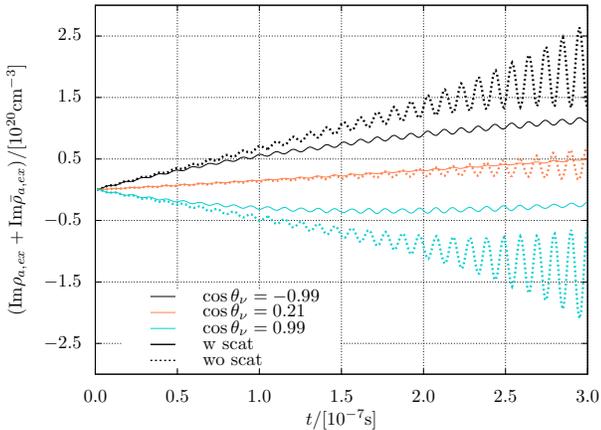}
    \caption{The time evolution of ${\rm Im}\rho_{a,ex}+{\rm Im}\bar{\rho}_{a,ex}$ at three different directions: $\cos \theta_\nu = -0.99$ (black), $0.21$ (orange), and $0.99$ (light blue), for a test of fast flavor conversion with scatterings. Color distinguishes angular directions. Solid and dashed lines represent the results with and without scatterings, respectively.}
    \label{time_evo_theta}
\end{figure}

To see the impact of scatterings on fast flavor conversion, we show the time evolution of off-diagonal component of $\rho_a$ and $\bar{\rho}_a$ in Fig.~\ref{time_evo_theta}. We pick up three different angular positions in the figure. This figure illustrates that both the oscillation amplitude and the growth rate in flavor conversion are suppressed by scatterings regardless of neutrino and anti-neutrinos. This may be attributed to the fact that the ELN angular distribution evolves to an isotropic distribution by scatterings (see Fig.~\ref{angle_e}), i.e., the amplitude of ELN crossing decreases with time. As discussed in \citet{sherwood2021}, the reduction of the amplitude of ELN crossing may account for the suppression. We reckon that there is another reason. As shown in Fig.~\ref{angle_dist}, the fast flavor conversion occurs vigorously around the angular point of ELN crossing. On the other hand, the crossing point is migrated by scatterings (see Fig.~\ref{angle_e}), that works to disperse the growth of flavor conversions into wide angles. This would result in suppressing flavor conversion, unless the flavor conversion grows substantially during the stagnation of the crossing. This suggests an interesting possibility that the effect of scatterings hinges on the cross section of scatterings, which is in line with results in \citet{martin2021} and
discussions in \citet{shalgar2021b}.

\begin{figure}[htbp]
    \includegraphics[width=\columnwidth]{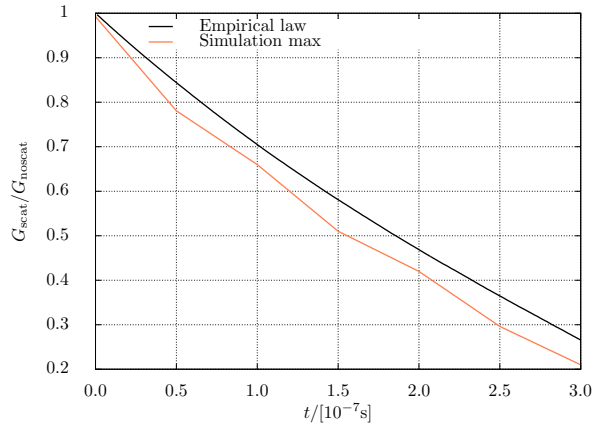}
    \caption{The time evolution of the growth rate ratio of fast flavor conversion between with and without scatterings. The ratio of $< 1$ indicates that the scattering suppresses fast flavor conversion. Orange line denotes the result computed by the time evolution of ${\rm Im}\rho_{ex}+{\rm Im}\bar{\rho}_{ex}$ from our QKE-MC simulation. The ratio computed from the empirical laws by \citet{morinaga2020} is shown in black line. See text for more detail.}
    \label{comp_growth}
\end{figure}

To strengthen the above argument, we compare the growth rate of fast flavor conversion obtained from the simulation to an empirical estimation proposed in our previous paper \citep{morinaga2020}. We estimate the growth rate from our simulation as follows. We take the peak amplitude of off-diagonal component of $\rho_a$ and $\bar{\rho}_a$ on each propagation direction (see, e.g., Fig.~\ref{time_evo_theta}). We compute the slope (growth rate) by computing the difference at adjacent peak points and divided by the difference of time, and then take the angular average; the result is plotted in black line of Fig.~\ref{comp_growth}. It should be noted that the growth rate is normalized by the one without scattering. Our result suggests that scatterings under the choice of the parameters reduce the growth of flavor conversion by $\sim$70\% at $t\sim3\times10^{-7}$~s.
The orange line in the same figure, on the other hand, portrays the growth rate obtained by an empirical formula,
\begin{eqnarray}
G \sim \sqrt{\left( \int_{\Delta\rho_a > 0} \Delta\rho_a d\cos{\theta_\nu} \right) \left(\int_{\Delta\rho_a < 0} \Delta\rho_a d\cos{\theta_\nu} \right)}, \ \ \ \ \ \ 
\end{eqnarray}
with $\Delta\rho_a\equiv\rho_{a,ee}-\bar{\rho}_{a,ee}$ (see \citet{morinaga2020}). We estimate the growth rate at each timestep by using $\Delta \rho_a$ obtained from the simulation. Although the two lines in Fig.~\ref{comp_growth} are slightly different from each other, the qualitative trend is essentially the same\footnote{The deviation is mainly due to the accuracy of the empirical relation; indeed, the exact growth rate is quantitatively different from it. See \citet{morinaga2020} for more detail.}. This result suggests that the growth of flavor conversion is determined mainly by the angular distribution of $\nu_e$ and $\bar{\nu}_e$ at each instantaneous time, and that the isotropized ELN angular distribution due to scatterings is the major factor of the suppression (see Fig.~\ref{angle_e}). It should also be stressed that the reasonable agreement between the two results suggests that our QKE-MC solver works well to this problem.

\begin{figure}[htbp]
 \centering
    \includegraphics[width=\columnwidth]{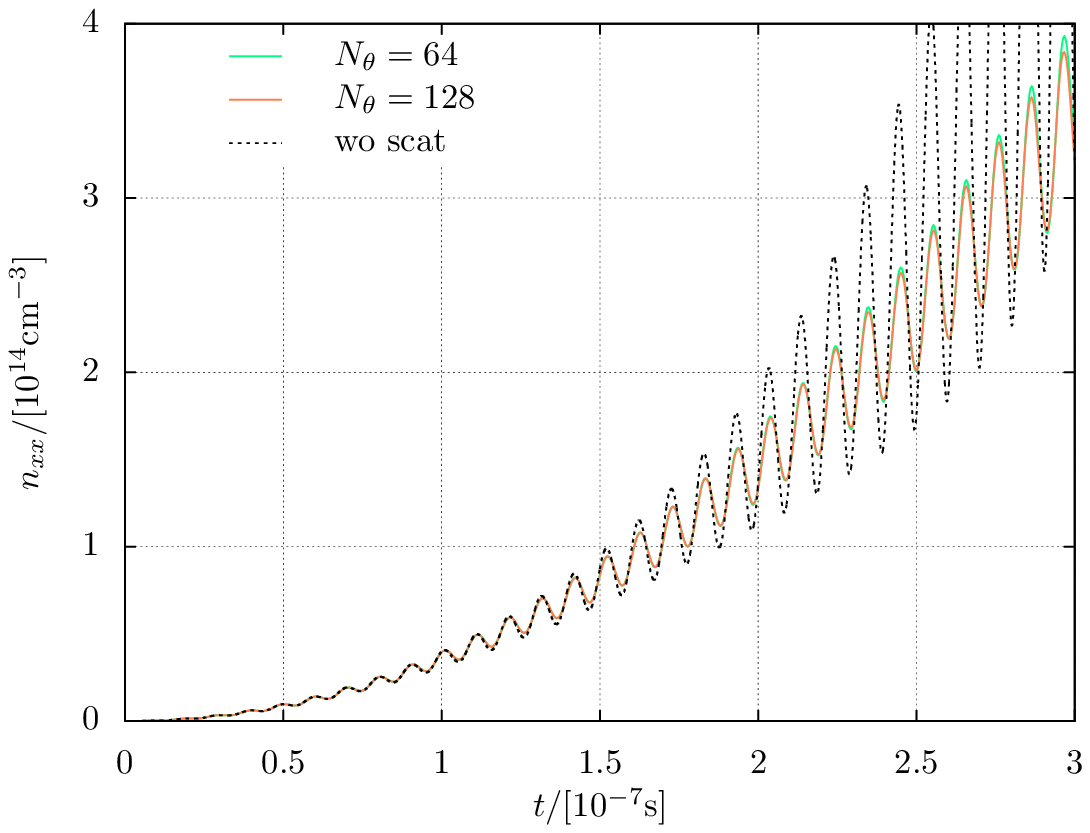} \\
    \includegraphics[width=\columnwidth]{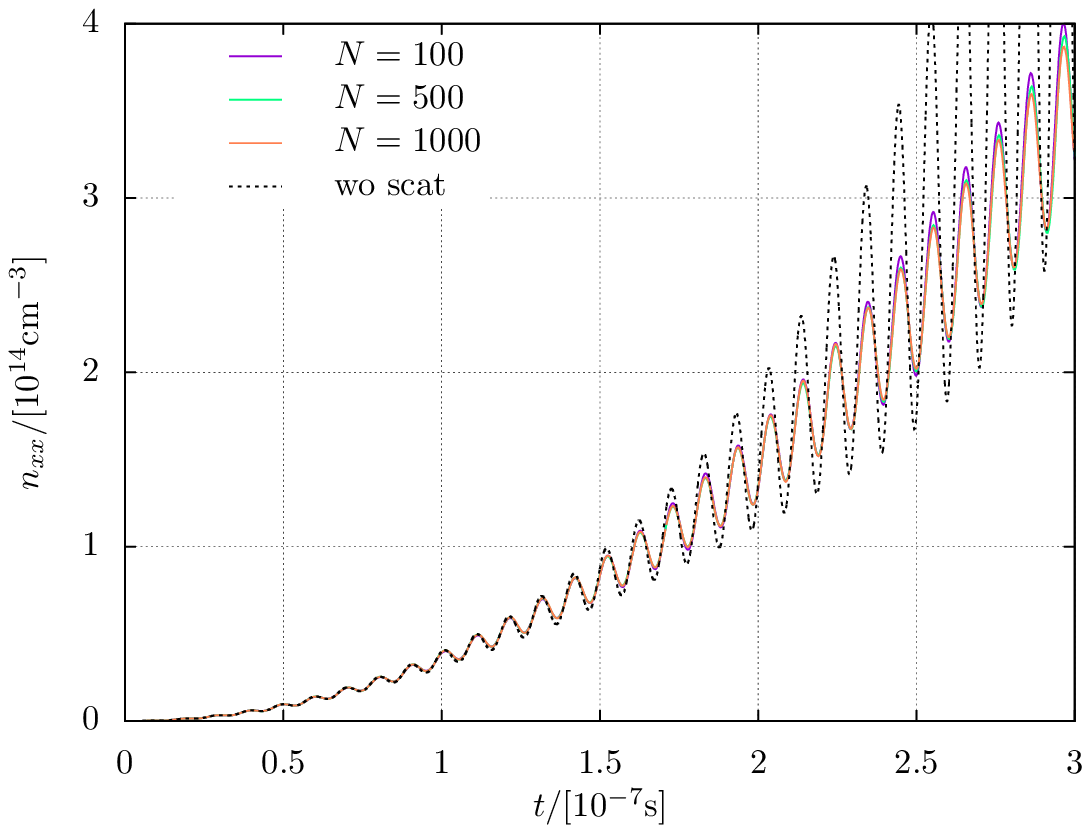} \\
    \includegraphics[width=\columnwidth]{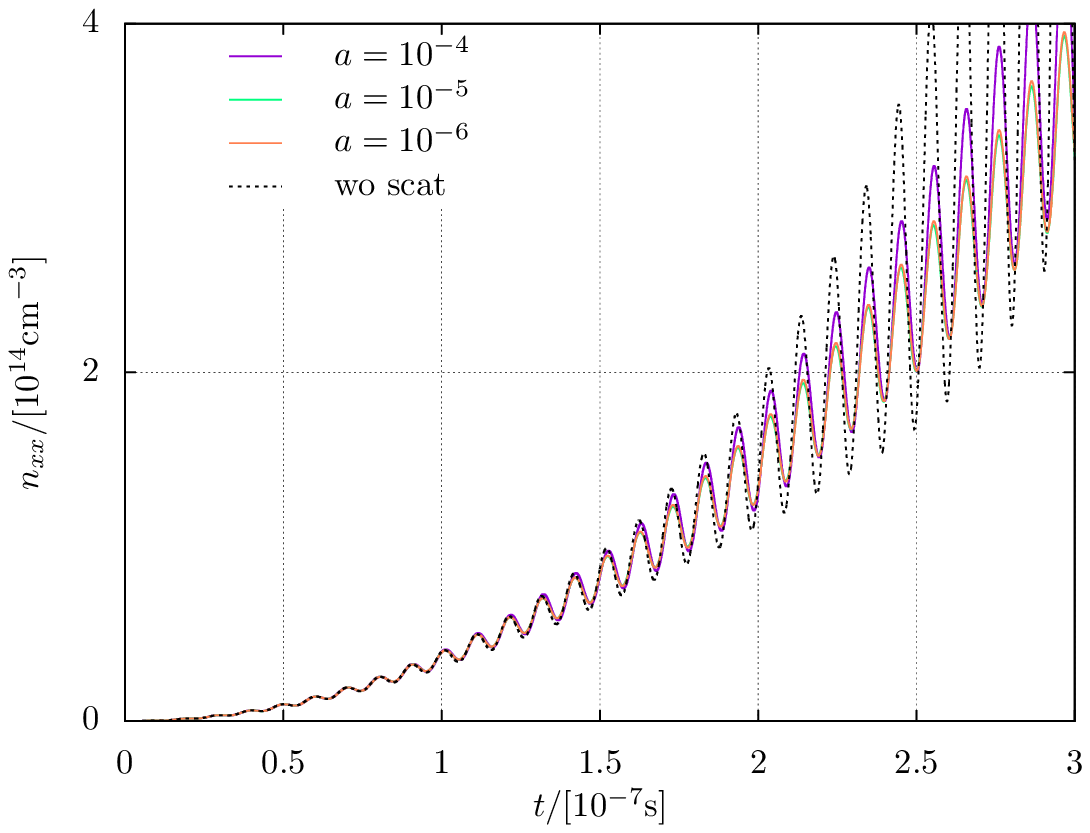} \\
    \caption{Resolution studies for the test of fast flavor conversion with scatterings (Sec.~\ref{ch7-3}). Top: the number of angular grids ($N_{\theta}$). Middle: the number of sample particles in each grid ($N$). Bottom: EMFP parameter ($a$).
    Color distinguishes different parameters for each resolution test.
    Green line corresponds to a reference calculation ($N_\theta=64$, $N=500$, $a=10^{-5}$).
    For comparison, the results without scatterings are also plotted with black dashed lines.}
    \label{resolution_check}
\end{figure}

Finally, we perform resolution studies.
In the top panel of Fig.~\ref{resolution_check}, we show the resolution dependence of $N_{\theta}$. The reference value is 64, but we check the case with 128. It should be mentioned that the number of MC particles is also increased twice from the reference in the case with $N_{\theta}=128$. This figure suggests that the difference between the two results is subtle. We also show the result without scatterings in the same figure, and the difference to the cases with scatterings is remarkable. Hence, the resolution of $N_{\theta}=64$ is sufficient to capture the impact of scatterings qualitatively. In the middle panel of Fig.~\ref{resolution_check}, we study the influence of statistical noise. For convenience, we use another variable, $N$, that is defined as the total number of MC particles divided by $N_{\theta}$, i.e., it represents the total number of MC particles on each angular mesh. We find that $N=500$ (reference value) is almost identical to the case with $N=1000$, indicating that the statistical noise in our reference model does not compromise our results. It should be mentioned that the statistical noise is also reduced by the EMFP method. To see the impact more quantitatively, we run two additional simulations with $a=10^{-4}$ and $10^{-6}$, and the results are displayed in the bottom panel of Fig.~\ref{resolution_check}. This figure suggests that the simulation with $a=10^{-5}$ (reference value) is capable of providing a physically accurate result. This convergence test strengthens the reliability and capability of our new modules. It should also be mentioned that we made a comparison to the group of \citet{shalgar2021b}, and confirmed that both simulations provide consistent results.

\section{Summary and discussions} \label{ch8}
Recent years have witnessed a number of theoretical indications that collective neutrino oscillations commonly occur in CCSN and BNSM environments. This suggests that quantum kinetic features as neutrino oscillations needs to be taken into account, that potentially impact the global dynamics, nucleosynthesis, and neutrino signals. The full QKE neutrino transport is capable of describing both the microscopic and macroscopic dynamics. Numerical simulations may be the most general ab-initio approach. Although a considerable amount of efforts has been devoted to numerical simulations, the physically consistent treatments of flavor conversion, transport, and matter collision are technically difficult. In this paper, we propose a Monte Carlo (MC) method, which is potentially capable of handling all elements self-consistently. To this end, we updated our classical MC neutrino transport solver \citep{kato2020} to that of QKE. It is a representative probabilistic and particle method, indicating that our numerical models will be complementary to those simulated by other deterministic and mesh-based approaches.

In this QKE-MC solver, we handle neutrino flavor conversions by embedding a flavor degree of freedom into each MC particle (see Sec.~\ref{subsec:transOsc}). With the upgrades, numerical treatments of collision terms, in particular for scattering processes, need to be changed (see Sec.~\ref{subsec:col}). New techniques are also developed to reduce statistical noise inherent in MC method. In this paper, we present a suite of code tests (see Secs.~\ref{ch4}~to~\ref{ch7}) to see the capability of each upgraded module by taking a cross-validation strategy, in which we reproduce the results reported in the previous literature. We show that our results are in good agreement with them for the tests of vacuum, matter, and collective oscillations. On the other hand, we did not reproduce the result of \citet{shalgar2021b} in tests of fast flavor conversions in cases without scatterings. We, hence, validated our results by applying another code (finite-difference scheme) to the same problem, and made a quantitative comparison, that provides consistent results with QKE-MC solver. In cases including scatterings, we validated our results by comparing an empirical (analytic) formula, and we showed that the QKE-MC solver is capable of providing physically accurate solutions.
It should also be mentioned that we and the authors of \citet{shalgar2021b} made an in-depth analysis on each simulation. We finally reached a consensus that the QKE-MC solver is capable of providing physically accurate solutions.

The present paper is the first report describing the essential philosophy, design, and implementation of QKE neutrino transport by a MC approach. The detailed investigations of non-linear dynamics entangling neutrino oscillation, transport, and collision term are not the focus of this work but the greatest motivation. One of our short-term targets is to reveal the effects of collision terms, in particular for scatterings, on fast flavor conversions. We note that the results of numerical simulations by different groups are still at odds each other \citep[see, e.g.,][]{capozzi2019,shalgar2021b,martin2021}. Since our numerical method is fundamentally different from theirs, the results will provide an unique insight into the source of controversy. We have already started the long-term simulations of such a system, and the results will be reported in a forthcoming paper.

\begin{acknowledgements}
We are grateful to Sherwood Richers for valuable comments and discussions. We also thank Irene Tamborra and Shashank Shalgar for carefully checking their simulation to resolve the discrepancy between us. C.K. is supported by Grant-in-Aid for Early-Career Scientists (No.20K14457) and Grant-in-Aid for Scientific Research on Innovative Areas (No.20H05240). T.M. is supported by JSPS Grant-in-Aid for JSPS Fellows (No. 19J21244) from the Ministry of Education, Culture, Sports, Science and Technology (MEXT), Japan.
\end{acknowledgements}

\bibliography{paper}


\end{document}